\documentclass[letterpaper,twocolumn,10pt]{article}
\makeatletter
\@namedef{ver@breakurl.sty}{}%
\@namedef{ver@cite.sty}{}%
\makeatother
\usepackage[hyphens]{url}
\usepackage{usenix}
\usepackage[sort,numbers]{natbib}

\date{}

\author{
{\rm Alejandro Cabrera Aldaya} \hspace{3em} {\rm Billy Bob Brumley}\\
Tampere University, Tampere, Finland\\
\{alejandro.cabreraaldaya,billy.brumley\}@tuni.fi
}

\usepackage[T1]{fontenc}
\usepackage{graphicx}
\usepackage{alltt}
\usepackage{xspace}
\usepackage{lipsum}
\usepackage{tcolorbox}
\tcbset{colback=white,boxsep=0mm}
\tcbuselibrary{theorems}

\newtcbtheorem[auto counter]{researchquestion}{RQ}{}{rq}
\makeatletter
\newcommand\tcb@cnt@researchquestionautorefname{RQ}
\makeatother

\DeclareMathOperator{\abs}{abs}
\newcommand{\code}[1]{\texttt{#1}\xspace}
\newcommand{\victimzero}{\code{x64\_\-vic\-tim\_\-0}}
\newcommand{\victimone}{\code{x64\_\-vic\-tim\_\-1}}
\newcommand{\HD}{\textsc{Hyper\-Degrade}\xspace}
\newcommand{\DG}{\textsc{Degrade}\xspace}
\newcommand{\NO}{\textsc{No\-Degrade}\xspace}
\newcommand{\CVE}[1]{\href{https://cve.mitre.org/cgi-bin/cvename.cgi?name=CVE-#1}{\mbox{CVE-#1}}}
\newcommand{\footurl}[1]{\footnote{\url{#1}}}
\newcommand{\rfc}[1]{\href{https://tools.ietf.org/html/rfc#1}{RFC #1}~\citep{rfc:#1}}
\newcommand{\fl}{\textsc{Flu\-sh+\allowbreak Re\-load}\xspace}
\newcommand{\ie}{i.e.,\xspace}
\newcommand{\eg}{e.g.,\xspace}

\newcommand{\clflush}{\code{clflush}}
\newcommand{\Paragraph}[1]{\smallbreak\noindent\textbf{#1.}}
\makeatletter
\let\PERCENTHACK\@percentchar
\makeatother
\newcommand{\RIO}{{\PERCENTHACK}r10}
\newcommand{\RDI}{{\PERCENTHACK}rdi}
\newcommand{\RSI}{{\PERCENTHACK}rsi}
\newcommand{\DIL}{{\PERCENTHACK}dil}
\newcommand{\RIP}{{\PERCENTHACK}rip}
\newcommand{\RCX}{{\PERCENTHACK}rcx}
\newcommand{\BBBPAD}[1]{\makebox[2.2em][r]{#1}}

\title{{HyperDegrade}: From {GHz} to {MHz} Effective {CPU} Frequencies}

\begin{document}

\maketitle

\begin{abstract}
Performance degradation techniques are an important complement
to side-channel attacks.
In this work, we propose \HD---a combination of previous approaches and
the use of simultaneous multithreading (SMT) architectures.
In addition to the new technique, we investigate the root causes
of performance degradation using cache eviction,
discovering a previously unknown slowdown origin.
The slowdown produced is significantly higher than previous approaches,
which translates into an increased time granularity for \fl attacks.
We evaluate \HD on different Intel microarchitectures,
yielding significant slowdowns that achieve, in select microbenchmark cases,
three orders of magnitude improvement over state-of-the-art.
To evaluate the efficacy of performance degradation in side-channel amplification,
we propose and evaluate leakage assessment metrics.
The results evidence that \HD increases time granularity without a meaningful impact on trace quality.
Additionally, we designed a fair experiment that compares three performance
degradation strategies when coupled with \fl from an attacker perspective.
We developed an attack on an unexploited vulnerability in OpenSSL
in which \HD excels---reducing by three times the number of required \fl traces to succeed.
Regarding cryptography contributions,
we revisit the recently proposed Raccoon attack on TLS-DH key exchanges,
demonstrating its application to other protocols.
Using \HD{}, we developed an end-to-end attack that shows
how a Raccoon-like attack can succeed with real data,
filling a missing gap from previous research.

 \end{abstract}
\section{Introduction} \label{sec:intro}
Side Channel Analysis (SCA), is a cryptanalytic technique that targets the
implementation of a cryptographic primitive rather than the formal mathematical
description. Microarchitecture attacks are an SCA subclass that focus on
vulnerabilities within the hardware implementation of an Instruction Set
Architecture (ISA). While more recent trends exploit speculation
\cite{DBLP:conf/uss/Lipp0G0HFHMKGYH18,DBLP:conf/sp/KocherHFGGHHLM019}, classical
trends exploit contention within different components and at various levels.
Specifically for our work, the most relevant is cache contention.

\citet{Percival05} and \citet{DBLP:conf/ctrsa/OsvikST06} pioneered access-driven
L1 data cache attacks in the mid 2000s, then \citet{DBLP:conf/ches/AciicmezBG10}
extended to the L1 instruction cache setting in 2010. Most of the threat models
considered only SMT architectures such as Intel's
HyperThreading (HT), where victim and spy processes naturally
execute in parallel. \citet{DBLP:conf/uss/YaromF14} removed this requirement
with their groundbreaking \fl technique utilizing cache line
flushing \cite{DBLP:conf/sp/GullaschBK11}, encompassing cross-core attacks in the
threat model by exploiting (inclusive) Last Level Cache (LLC) contention.

In this work, we examine the following Research Questions (RQ).

\begin{researchquestion}{Research Question \ref*{rq:hd}}{hd}
With respect to SMT architectures,
are CPU topology and affinity factors in performance degradation attacks?
\end{researchquestion}
\citet{DBLP:conf/acsac/AllanBFPY16} proposed \DG as a general performance
degradation technique, but mainly as a companion to \fl attacks. They identify
hot spots in code and repeatedly flush to slow down victims---in the \fl case,
with the main goal of amplifying trace granularity. \citet{DBLP:conf/uss/GarciaB17}
proposed an alternate framework for hot spot identification.
We explore \autoref{rq:hd} in \autoref{sec:concept} to understand what role
physical and logical cores in SMT architectures play in performance degradation.
Along the way, we discover the root cause of \DG which we subsequently amplify.
This leads to our novel \HD technique, and \autoref{sec:perf} shows its efficacy,
with slowdown factors in select microbenchmark cases remarkably exceeding three
orders of magnitude.

\begin{researchquestion}{Research Question \ref*{rq:nicv}}{nicv}
Does performance degradation lead to \fl traces with statistically more info.\ leakage?
\end{researchquestion}
Nowadays, \fl coupled with \DG is a standard offensive technique for academic
research. While both \citet{DBLP:conf/acsac/AllanBFPY16} and
\citet{DBLP:conf/uss/GarciaB17} give convincing use-case specific motivation for
why \DG is useful, neither actually show the information-theoretic advantage of
\DG. \autoref{sec:eval} closes this gap and partially answers \autoref{rq:nicv} by
utilizing an existing SCA metric to demonstrate the efficacy of \DG as an SCA
trace amplification technique.
We then extend our analysis to
our \HD technique to resolve \autoref{rq:nicv}. At a high level, it shows \HD
leads to slightly noisier individual measurements yet positively
disproportionate trace granularity.

\begin{researchquestion}{Research Question \ref*{rq:realHD}}{realHD}
Can \HD reduce adversary effort when attacking crypto implementations?
\end{researchquestion}
\autoref{rq:nicv} compared \HD with previous approaches from
a theoretical point of view.
In \autoref{sec:attack} we compare the three approaches from an \emph{applied} perspective,
showing a clear advantage for \HD over the others.

\begin{researchquestion}{Research Question \ref*{rq:raccoon}}{raccoon}
Can a Raccoon attack (variant) succeed with real data?
\end{researchquestion}
\citet{tmp:raccoon} recently proposed the Raccoon attack (e.g.\ \CVE{2020-1968}),
a timing attack targeting recovery of TLS 1.2 session keys by exploiting DH
key-dependent padding logic. Yet the authors only model the SCA data and
abstract away the protocol messages. \autoref{sec:attack} answers
\autoref{rq:raccoon} by developing a microarchitecture timing attack variant of
Raccoon, built upon \fl and our new \HD technique. Our end-to-end attack uses
real SCA traces and real protocol (CMS) messages to recover session keys,
leading to loss of confidentiality.
We conclude in \autoref{sec:conclusion}.
\section{Background} \label{sec:background}
\subsection{Memory Hierarchy}
Fast memory is expensive, therefore computer system designers use faster yet
smaller caches of slower yet larger main memory to benefit from locality
without a huge price increase.
A modern microprocessor has several caches (L1, L2, LLC) forming a cache
hierarchy \cite[Sect.\ 8.1.2]{DBLP:series/txcs/Page09},
the L1 being the fastest one but smaller and tightly coupled to the processor.
Caches are organized in cache lines of fixed size (\eg 64 bytes).
Two L1 caches typically exist, one for storing instructions and the other for data.
Regarding this work, we are mainly interested in the L1 instruction cache and remaining levels.

When the processor needs to fetch some data (or instructions) from memory,
it first checks if they are already cached in the L1.
If the desired cache line is in the L1, a \emph{cache hit} occurs and
the processor gets the required data quickly.
On the contrary if it is not in the L1, a \emph{cache miss} occurs and the processor tries to fetch it from
the next, slower, cache levels or in the worst case, from main memory.
When gathering data, the processor caches it to reduce latency in future loads of the same data,
backed by the principle of locality \cite[Sect.\ 8.1.5]{DBLP:series/txcs/Page09}.

\subsection{Performance Degradation}
In contrast to generic CPU monopolization methods like the ``cheat'' attack by \citet{DBLP:conf/uss/TsafrirEF07}
that exploit the OS process scheduler,
several works have addressed the problem of degrading the performance of
a victim using microarchitecture components
\citep{DBLP:conf/micro/GrunwaldG02,DBLP:conf/hpca/HasanJVB05,DBLP:conf/uss/MoscibrodaM07,DBLP:conf/uss/GrussSM15}.
However, in most cases it is not clear whether SCA-based attackers gain benefits
from the proposed techniques.

On the other hand, \citet{DBLP:conf/acsac/AllanBFPY16} proposed a cache-eviction based
performance degradation technique that enhances \fl attack SCA signals (traces).
This method has been widely employed in previous works to mount SCA attacks on cryptography implementations.
For instance RSA \citep{DBLP:conf/ches/BernsteinBGBHLV17},
ECDSA \citep{DBLP:conf/ccs/AranhaN0TY20},
DSA \citep{DBLP:conf/ccs/GarciaBY16},
SM2 \citep{DBLP:conf/acsac/TuveriHGB18},
AES \citep{DBLP:conf/sp/CohneyKPGHRY20}, and
ECDH \citep{DBLP:conf/ccs/GenkinVY17}.

The performance degradation strategy proposed by \citet{DBLP:conf/acsac/AllanBFPY16},
\DG from now on, consists of an attacker process that causes cache contention by
continuously issuing \clflush instructions.
It is an unprivileged instruction that receives a virtual memory address as an operand
and evicts the corresponding cache line from the entire memory hierarchy \citep{intel_sdm_vol3}.

This attack applies to shared library scenarios, which are common in many OSs.
This allows an attacker to load the same library used by the victim and receive a virtual address
that will point to the same physical address, thus, same cache line.
Therefore, if the attacker evicts said cache line from the cache,
when the victim accesses it (\eg executes the code contained within it),
a cache miss will result, thus the microprocessor must fetch the content from slower main memory.

\subsection{Leakage Assessment} \label{sec:leakage_assess}
Pearson's correlation coefficient, Welch's T-test, Test Vector Leakage
Assessment (TVLA), and Normalized Inter-Class Variance (NICV) are established
statistical tools in the SCA field. Leakage assessment leverages these
statistical tools to identify leakage in procured traces for SCA. A short
summary follows.

Pearson's correlation coefficient measures the linear similarity between two random variables.
It is generally useful for leakage assessment \cite[Sect.\ 3.5]{DBLP:conf/fc/CoronKN00} and
Point of Interest (POI) identification within traces, for example in template
attacks \cite{DBLP:conf/ches/ChariRR02} or used directly in Correlation Power
Analysis (CPA) \cite{DBLP:conf/ches/BrierCO04}. POIs are the subset of points
in an SCA trace that leak sensitive information.

Welch's T-test is a statistical measure to determine if two sample sets were
drawn from populations with similar means.
\citet{Goodwill11p:a} proposed TVLA that utilizes the T-test for leakage
assessment by comparing sets of traces with fixed vs.\ random cryptographic keys
and data.

Lastly, \citet{2014:NICV} propose NICV for leakage assessment. It is an ANalysis
Of VAriance (ANOVA) F-test, a statistical measure to determine if a number of
sample sets were drawn from populations with similar variances.

\subsection{Key Agreement and SCA} \label{sec:raccoon}
\citet{tmp:raccoon} recently proposed the Raccoon attack that exploits a
spe\-ci\-fi\-ca\-tion-le\-vel weakness in protocols that utilize Diffie-Hellman key
exchange. The key insight is that some standards, including TLS 1.2 and below,
dictate stripping leading zero bytes from the shared DH key (session key, or
pre-master secret in TLS nomenclature). This introduces an
SCA attack vector since, at a low level, this behavior trickles down to several
measurable time differences in components like compression functions for hash
functions. In fixed DH public key scenarios, an
attacker observes one TLS handshake (the target) then repeatedly queries the
victim using a large number of TLS handshakes with chosen inputs. Detecting
shorter session keys through timing differences, the authors use these inputs to
construct a lattice problem to recover the target session key, hence
compromising confidentiality for the target TLS session.
\section{HyperDegrade: Concept} \label{sec:concept}
The objective of \HD is to improve performance degradation offered by \DG when targeting a victim process,
resulting in enhanced SCA traces when coupled with a \fl attack.
Under a classical \DG attack,
the degrading process continuously evicts a cache line from the cache hierarchy,
forcing the microprocessor to fetch the cache line from main memory when the victim needs it.

It would be interesting to evaluate the efficacy of the \DG strategy, seeking avenues for improvement.
The root cause of \DG as presented in \citep{DBLP:conf/acsac/AllanBFPY16} is the cache will produce more misses
during victim execution---we present novel results on this later.
Therefore, the cache miss to executed instructions ratio is a reasonable metric to evaluate its performance.

For this task, we developed a proof-of-concept victim that executes custom code located in a shared library.
This harness receives as input a cache line index, then executes a tight loop in said cache line several times.
\autoref{fig:concept_code} shows the code snippet of this loop at the left,
and one cache line disassembled code at the right.

For our experiments the number of iterations executed is $2^{16}$ (defined by \code{rsi}).
Therefore, we expect the number of instructions executed in the selected cache line is about 1M.
Under normal circumstances, every time the processor needs to fetch this code from memory,
the L1 cache should serve it very quickly.

\begin{figure}
\scriptsize
\begin{alltt}
                            5000:   add    $0x1,\RSI
                            5004:   sub    $0x1,\RSI
.p2align 12                 5008:   add    $0x1,\RSI
L0:                         500c:   sub    $0x1,\RSI
.rept 64                    5010:   add    $0x1,\RSI
    .rept 6                 5014:   sub    $0x1,\RSI
        add $1, \RSI        5018:   add    $0x1,\RSI
        sub $1, \RSI        501c:   sub    $0x1,\RSI
    .endr                   5020:   add    $0x1,\RSI
    add $1, \RSI            5024:   sub    $0x1,\RSI
    sub $2, \RSI            5028:   add    $0x1,\RSI
    jz END                  502c:   sub    $0x1,\RSI
    jmp *\RDI ; L0          5030:   add    $0x1,\RSI
    .p2align 6              5034:   sub    $0x2,\RSI
.endr                       5038:   je     6000 <END>
                            503e:   jmpq   *\RDI      ; L0
\end{alltt}
\vspace{-4ex}
\caption{Victim single cache line loop (code and disasm.).}
\label{fig:concept_code}
\end{figure}

\subsection{Degrade Revisited}\label{sec:degrade_revisited}
On the \DG attacker side, we developed a degrading process that loads the same shared library and continuously evicts
the victim executed cache line using \clflush.
We use the Linux \code{perf} tool to gather statistics about victim execution under a \DG attack.
For this task, we used the \code{perf} (commit \code{13311e74}) FIFO-based performance counters control to sync their
sampling with the victim and degrade processes.
\code{perf} uses two FIFOs for this task, one for enabling/disabling the performance counters and another
for giving ACKs.
The sync procedure in our measurement tooling is the following:

\begin{enumerate}
    \item The degrade process executes and it blocks until receiving an ACK packet from \code{perf} using FIFO \code{A}.
    \item \code{perf} executes with counters disabled (``-D -1'' option), using FIFO \code{C} for control and \code{A} for ACKs.
          Then it runs \code{taskset} that executes the victim pinned to a specific core.
    \item The victim enables the counters by writing to \code{C},
          then it blocks until it receives an ACK from the degrade process using another FIFO.
    \item When \code{perf} receives the enable counters command, it sends an ACK using \code{A} to the degrade process.
          When the latter receives the ACK, it forwards it to the victim.
          When the victim receives this packet, it starts executing its main loop (\autoref{fig:concept_code}).
    \item Once the victim finishes, it disables the counters in \code{perf}.
\end{enumerate}

This procedure considerably reduces measurement tooling overhead, but some remains.
The \NO strategy does not use a degrade process,
however we used a dummy process that follows the FIFO logic to unify the sync procedure among experiments.
We repeated each experiment 100 times, gathering the average and relative standard deviation.
In all reported cases the latter was less than 4\%, therefore we used the average for our analysis.
We recorded the number of L1 instruction cache misses and the number of instructions retired by the microprocessor.
For these experiments, we used the environment setup Coffee Lake detailed in \autoref{tab:machines}.

We collected data while the victim was running standalone (\ie \NO strategy) and while it was under \DG effect.
\autoref{tab:perf1} shows the results for each \code{perf} event.
The number of retired instructions is roughly the same between both experiments,
where the difference from expected (1M) is likely due to the measurement tooling overhead.
Nevertheless, the number of L1 instruction cache misses was 4k for the \NO test
and 33k for \DG.
However, 33k is still far below one cache-miss per executed instruction (1M).

\begin{table}
    \caption{\NO and \DG statistics.}
    \label{tab:perf1}
    \centering
    \resizebox{1.0\linewidth}{!}{%
        \begin{tabular}{lrr} \hline
            Parameter                    &  \NO   &   \DG  \\ \hline
            \code{inst\_retired.any}     &  1.5M  &   1.5M \\
            \code{L1-icache-load-misses} & 4,115  & 33,785 \\ \hline
        \end{tabular}
    }
\end{table}

\subsection{The HyperDegrade Technique}\label{sec:hd_concept}
In order to increase the performance impact of \DG,
we attempt to maximize the number of cache misses.
For this task we made the hypothesis that in an SMT architecture,
if the degrade process is pinned to the victim's sibling core,
then the number of cache misses will increase.

\begin{figure}
\includegraphics[width=\linewidth]{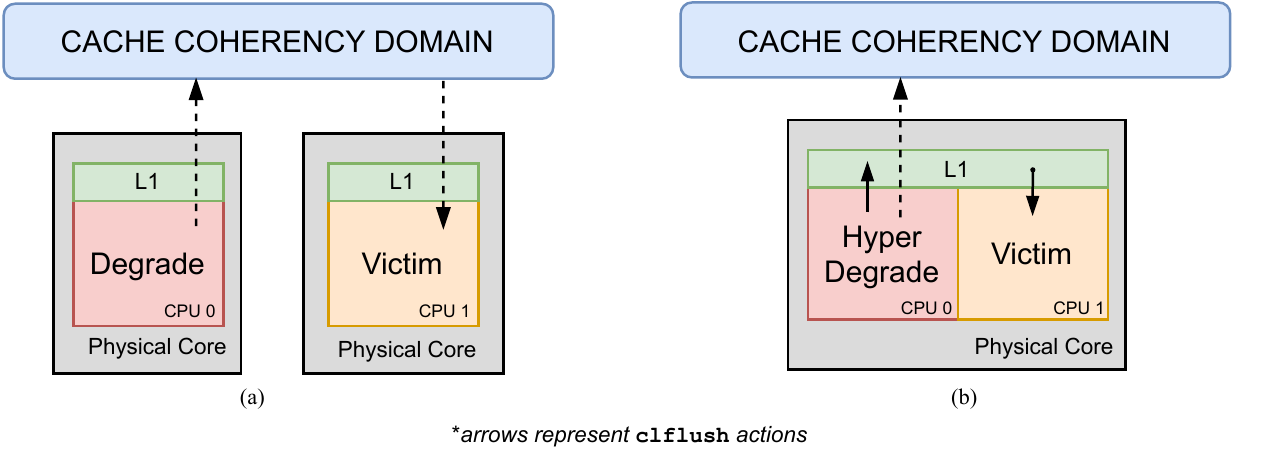}
\vspace{-4ex}
\caption{\DG vs \HD from \clflush perspective.}
\label{fig:concept}
\end{figure}

According to an expired patent from Intel concerning \clflush \citep{patent_clflush},
the microarchitectural implementation of this instruction in the ISA
distinguishes if the flushed cache line is already present in the L1 or not.
While it is not explicitly stated in that document as there is no latency analysis,
it is our belief that the flushed cache line would be evicted from the L1 before others caches,
\eg due to the proximity wrt., for instance, the LLC controller.
\autoref{fig:concept} illustrates this idea,
where the arrows represent \clflush actions and the dashed ones are slower than the others.

Following this hypothesis, we present \HD as a cache-evicting degrade strategy that runs in the
victim sibling core in a microarchitecture with SMT support.
From an architecture perspective it does the same task as \DG, but in the same physical core as the victim.
However, the behavior at the microarchitecture level is quite different because, if our hypothesis is correct,
it should produce more cache misses due to the local proximity of the L1.
To support this claim, we repeated the previous experiment while pinning the degrade process to the victim sibling core.

\autoref{tab:perf_hd} shows the results of \HD in comparison with the previous experiment.
Note that with \HD there are about 33x cache misses%
\footnote{after subtracting \NO cache misses to remove non-targeted code activity}
than with \DG, translating to a considerable increase
in the number of CPU cycles the processor spends executing the victim.
At the same time, the number of observed cache misses increased considerably, approaching the desired rate.
This result, while not infallible proof, supports our hypothesis that sharing the L1 with the victim process
should produce higher performance degradation.

\begin{table}
    \caption{\HD improvement.}
    \label{tab:perf_hd}
    \centering
    \resizebox{1.0\linewidth}{!}{%
        \begin{tabular}{lrrr} \hline
            Parameter                    & \NO        &    \DG      &    \HD      \\ \hline
            \code{inst\_retired.any}     &      1.5M  &       1.5M  &        1.5M \\
            \code{L1-icache-load-misses} &     4,115  &     33,785  &     992,074 \\
            \code{cycles}                & 1,252,211  & 12,935,389  & 504,395,314 \\  \hline
            \code{machine\_clears.smc}   &      $<1$  &     28,375  &    983,348 \\ \hline
        \end{tabular}
    }
\end{table}

On the other hand, note the number of CPU cycles increases by a higher factor (43x), which leads us to suspect there
could be another player that is influencing the performance degradation; further research is needed.
After repeating the experiment for several \code{perf} parameters,
we found an interesting performance counter that helps explain this behavior.

It is the number of \emph{machine clears} produced by \emph{self-modifying code} or SMC (\code{machine\_clears.smc}).
According to Intel, a \emph{machine clear or nuke} causes the entire pipeline
to be cleared, thus producing a \emph{severe performance penalty} \citep[19-112]{intel_sdm_vol3}.

Regarding the SMC classification of the machine clear,
when the attacker evicts a cache line, it invalidates a cache line from the victim L1 instruction cache.
This might be detected by the microprocessor as an SMC event.

The machine clears flush the pipeline, forcing the victim to re-fetch some instructions from memory,
thus increasing the number of L1 cache misses due to the degrade process action.
Therefore, it amplifies the effect produced by a cache miss, because sometimes the same instructions are fetched more than once.

Moreover, this analysis reveals an unknown performance degradation root cause of both \DG and \HD,
thus complementing the original research on \DG in \citep{DBLP:conf/acsac/AllanBFPY16}.
The performance degradation occurs due to an increased number of cache misses and due to increased machine clears,
where the latter is evidenced by the significant increase from zero (\NO) to 28k (\DG).
Likewise, \HD increases the number of cache misses and machine clears, thus, further amplifying the performance
degradation produced by \DG.
This demonstrates that the topology of the microprocessor and the affinity of
the degrade process have significant influence in
the performance degradation impact, answering \autoref{rq:hd}.

We identified SMC machine clears as an additional root cause for both \DG and \HD,
however, there could be others.
In this regard, we highlight that our root cause analysis,
albeit sound, is not complete.
Moreover, achieving such completeness is challenging due to the undocumented nature
of the microarchitecture, providing an interesting research direction for continued research.
Indeed, in concurrent work, \citet{tmp:RagabBBG21} analyze machine clears in the context of transient execution.

\Paragraph{Contention test and pure SMC scenario}
For the sake of completeness, we compared the CPU cycles employed
by different experiments using the previous setup.
However, in this case, we vary the number
of iterations in the tight loop over a single cache line.
We ranged this value in the set $\{2^{16}, 2^{17},.., 2^{25}\}$.
Therefore, the number of executed instructions by the victim
will be $\text{\code{victim\_num\_inst}} = 16 \times \text{\code{num\_iter}}$.

This comparison involves five experiments: one for each degrade strategy,
plus a contention test and a pure SMC scenario (presented later).
The contention test is equivalent to \HD{}; however, this time
the \clflush instruction will flush a cache line \emph{not used} by the victim.
This test allows to evaluate the performance impact of co-locating a degrade process
while it does not modify the victim's cache state.

\begin{figure}
\includegraphics[width=\linewidth]{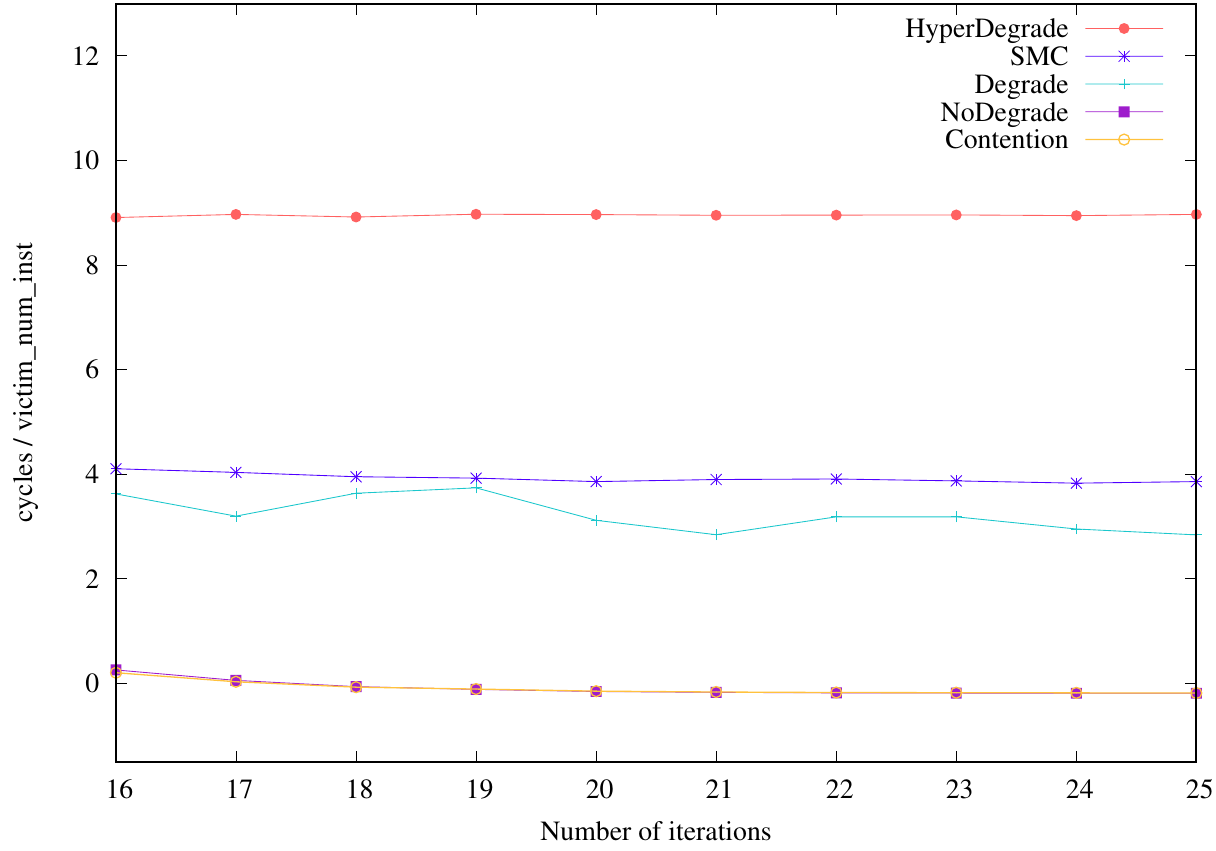}
\vspace{-4ex}
\caption{Experiments performance comparison ($\log_2$-scale).}
\label{fig:exp_cmp}
\end{figure}

\autoref{fig:exp_cmp} visualizes the results of this comparison,
where both axes are $\log_2$-scaled.
The $y$-axis represents the ratio cycles per \code{victim\_num\_inst}.
It can be appreciated that the contention test and \NO have very similar performance behavior
(\ie their curves overlap at the bottom).
Hence, the performance degradation of a \HD process will only be effective
if it flushes a victim cache line, whereas additional resource contention
related to executing the \clflush instruction in a sibling core can be neglected.

We included a pure SMC scenario as an additional degrade strategy.
\autoref{fig:pure_smc_code} illustrates the degrade process core.
This code continuously triggers machine clears due to its self-modifying code behavior.
Its position in \autoref{fig:exp_cmp} shows it has about the same performance degrading
power as \DG.
However, this pure SMC alternative does not depend on shared memory between the victim and
degrade processes, thus the presence of---and finding---hot
cachelines \cite{DBLP:conf/acsac/AllanBFPY16} is not a requirement.

\begin{figure}
\scriptsize
\begin{alltt}
    mov  $0x40, \DIL     ; 0x40 is part of the opcode at L0
    lea  L0(\RIP), \RCX  ; rcx points to L0
L0: mov  \DIL, (\RCX)
    jmp  L0
\end{alltt}
\vspace{-4ex}
\caption{Main SMC degrade process instructions.}
\label{fig:pure_smc_code}
\end{figure}

\Paragraph{Limitations}
\HD offers a significant slowdown wrt.\ previous performance degradation strategies.
On the other hand, it is tightly coupled to SMT architectures because it requires
physical core co-location with the victim process.
Therefore, it is only applicable to microprocessors with this feature.
In this regard, \HD has the same limitation as previous works that exploit
SMT \citep{Percival05,DBLP:conf/ches/AciicmezBG10,DBLP:conf/ches/YaromGH16,DBLP:conf/uss/GrasRBG18,DBLP:conf/sp/AldayaBHGT19,DBLP:conf/ccs/BhattacharyyaSN19}.
SCA attacks enabled by SMT often have no target shared library requirement, which is a hard requirement for \fl to move to cross-core application scenarios.
For example, neither the L1 dcache spy \cite{Percival05} nor the L1 icache spy
\cite{DBLP:conf/ches/AciicmezBG10} require victims utilizing any form of shared
memory on SMT architectures. Yet \HD retains this shared library requirement,
since our applet is based on \clflush to induce the relevant microarchitecture
events.
However, since SMT is a common feature in modern microarchitectures and shared
libraries are even more common, \HD is another tool on the attacker's belt for
performing \fl attacks.
\section{HyperDegrade: Performance} \label{sec:perf}
With our \HD applet from \autoref{sec:concept}, the goal of this section is to
evaluate the efficacy of \HD as a technique to degrade the performance of victim
applications that link against shared libraries. \autoref{sec:eval} will later
explore the use of \HD in SCA, but here we focus purely on the slowdown effect.
Applied as such in isolation, \HD is useful to effectively monopolize the CPU
comparative to the victim, and also increase the CPU time billed to the victim
for the same computations performed by the victim.

\citet[Sect.\ 4]{DBLP:conf/acsac/AllanBFPY16} use the SPEC 2006CPU benchmark suite,
specifically 29 individual benchmark applications, to establish the efficacy of
their \DG technique as a performance degradation mechanism. In our work, we
choose a different suite motivated from several directions.

First, unfortunately SPEC benchmarks are not free and open-source software
(FOSS). In the interest of Open Science, we instead utilize the BEEBS benchmark
suite by \citet{DBLP:journals/corr/PallisterHB13,DBLP:journals/cj/PallisterHB15}
which is freely available\footurl{https://github.com/mageec/beebs}. The original
intention of BEEBS is microbenchmarking of typical embedded applications
(sometimes representative) to facilitate device power consumption measurements.
Nevertheless, it suits our purposes remarkably.

These 77 benchmark applications also differ in the fact that they are not built
with debug symbols, which is required to apply the \citet{DBLP:conf/acsac/AllanBFPY16} methodology.
While debug symbols themselves should not affect application performance, they
often require less aggressive compiler optimizations that, in the end, result in
less efficient binaries which might paint an unrealistic picture for performance
degradation techniques outside of research environments.

We used the BEEBS benchmark suite off-the-shelf, with one minor modification. By
default, BEEBS statically links the individual benchmark libraries whereas \HD
(and originally \DG) target shared libraries. Hence, we added a new option to
additionally compile each benchmark as a shared library and dynamically link the
benchmark application against it.

\subsection{Experiment}
Before presenting and discussing the empirical results, we first describe our
experiment environment. Since \HD targets HT architectures
specifically, we chose four consecutive chip generations, all featuring HT.
\autoref{tab:machines} gives an overview, from older to younger
models.

\begin{table}
\caption{Various SMT architectures used in our experiments.}
\label{tab:machines}
\centering
\resizebox{1.0\linewidth}{!}{%
\begin{tabular}{lllrl} \hline
Family & Model & Base  & Cores / & Details \\
       &       & Freq. & Threads &         \\ \hline
Skylake & i7-6700 & 3.4 GHz & 4 / 8 & Ubuntu 18, 32 GB RAM \\
Kaby Lake & i7-7700HQ & 2.8 GHz & 4 / 8 & Ubuntu 20, 32 GB RAM \\
Coffee Lake & i7-9850H & 2.6 GHz & 6 / 12 & Ubuntu 18, 32 GB RAM \\
Whiskey Lake & i7-8665UE & 1.7 GHz & 4 / 8 & Ubuntu 20, 16 GB RAM \\ \hline
\end{tabular}
}
\end{table}

Our experiment consists of the following steps. We used the \texttt{perf}
utility to definitively measure performance, including clock cycle count. In an
initial profiling step, we exhaustively search (guided by \texttt{perf} metrics)
for the most efficient cache line to target during eviction.
We then run three different tests: a baseline \NO, classical \DG,
and our \HD from \autoref{sec:concept}. Each test that involves degradation
profiles for the target cache line independently: i.e.\ the target cache line
for \DG is perhaps not the same as \HD. We then iterate each test to gather
statistics, then repeat for all 77 BEEBS benchmarks, and furthermore across the
four target architectures. We used the \texttt{taskset} utility to pin to
separate physical cores in the \DG case, and same physical core in the \HD case.

\subsection{Results}
\begin{figure}
\includegraphics[width=\linewidth]{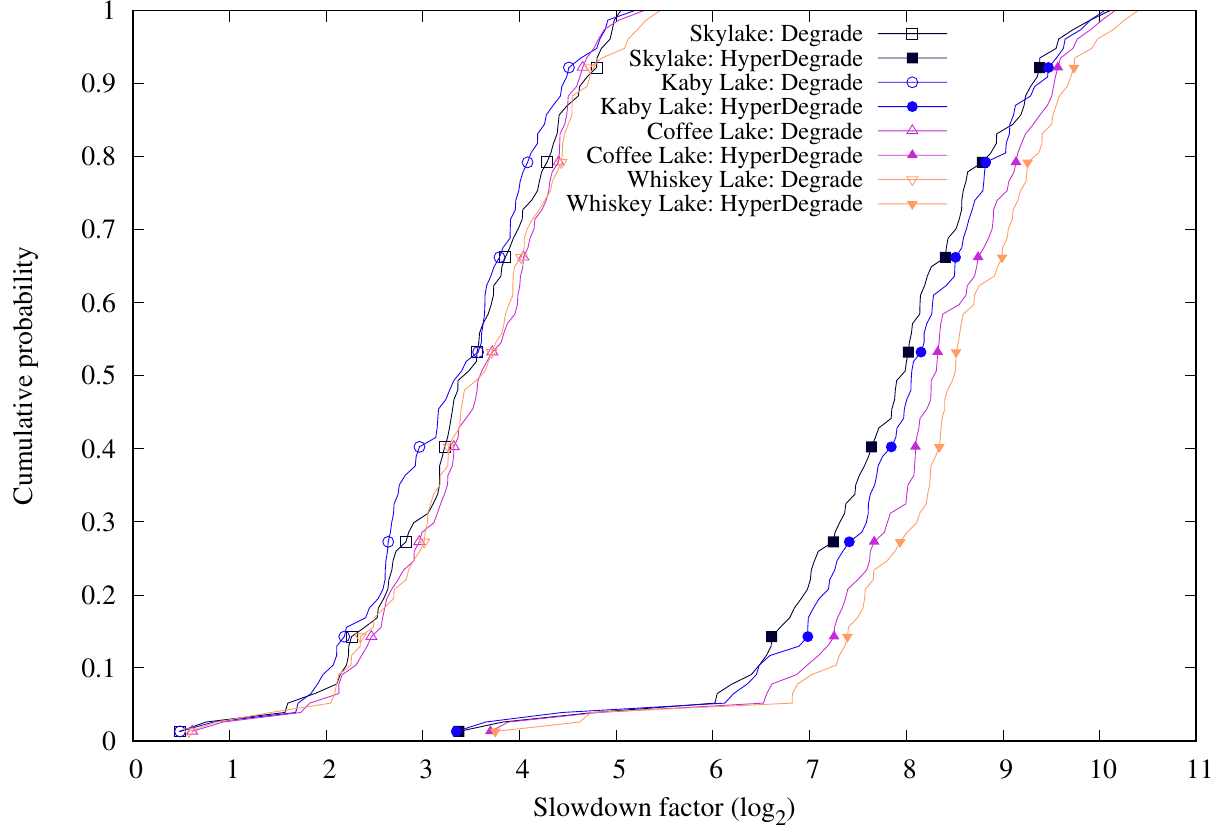}
\vspace{-4ex}
\caption{Distributions (computed from \autoref{tab:beebs2} and \autoref{tab:beebs3})
for different performance degradation strategies targeting BEEBS shared library
benchmarks, across architectures. Note the $x$-axis is logarithmic ($\log_2$).}
\label{fig:slowdown}
\end{figure}

While \autoref{tab:beebs2} and \autoref{tab:beebs3} contain the full statistics
per architecture, strategy, and BEEBS microbenchmark,
\autoref{tab:beebs1} and
\autoref{fig:slowdown} provide high level overviews of the aggregate data.
\autoref{tab:beebs1} shows the efficacy of \HD over classical \DG is striking,
with median slowdown factors ranging from 254 to 364, and maximum slowdown
factors ranging from 1060 to 1349. These maximum slowdowns are what our title
alludes to---for example, in the Skylake i7-6700 case (maximum), reducing the
3.4 GHz base frequency to a 3.1 MHz effective frequency when observed
from the victim application perspective.

\begin{table}
\caption{Statistics (aggregated from \autoref{tab:beebs2} and \autoref{tab:beebs3})
for different performance degradation strategies targeting BEEBS shared library
benchmarks, across architectures.}
\label{tab:beebs1}
\centering
\resizebox{1.0\linewidth}{!}{%
\begin{tabular}{llrrrrr} \hline
Family & Method & Median & Min & Max & Mean & Stdev \\ \hline
Skylake & \DG & 11.1 & 1.4 & 33.1 & 13.1 & 8.0 \\
Skylake & \HD & 254.0 & 10.4 & 1101.9 & 306.3 & 226.7 \\
Kaby Lake & \DG & 10.6 & 1.4 & 36.5 & 12.0 & 7.5 \\
Kaby Lake & \HD & 266.4 & 10.2 & 1060.1 & 330.6 & 229.0 \\
Coffee Lake & \DG & 12.2 & 1.5 & 39.0 & 14.0 & 7.9 \\
Coffee Lake & \HD & 317.5 & 13.0 & 1143.7 & 382.5 & 246.9 \\
Whiskey Lake & \DG & 12.5 & 1.5 & 43.9 & 14.4 & 9.2 \\
Whiskey Lake & \HD & 364.3 & 13.5 & 1349.3 & 435.8 & 280.9 \\
 \hline
\end{tabular}
}
\end{table}

\autoref{fig:slowdown} visualizes the aggregate statistics from
\autoref{tab:beebs2} and \autoref{tab:beebs3}. Due to the magnitude of the
slowdowns, the $x$-axis is logarithmic. Please note these data points
are for identifying general trends; the location of individual points within
separate distributions (i.e.\ different benchmarks) may vary.

Finally, \autoref{tab:parsec1} and \autoref{tab:parsec2} for the PARSEC
\cite{DBLP:conf/IEEEpact/BieniaKSL08} macrobenchmark suite are analogous to
\autoref{tab:beebs2} and \autoref{tab:beebs3} for the BEEBS microbenchmarks. In
this case, the slowdowns have a noticeably smaller magnitude. We attribute the
difference to benchmarking goals. While BEEBS microbenchmarks are typically
CPU-bound and capable of running on bare metal, that is not the case for PARSEC
macrobenchmarks where the focus is parallelism. The combined results from the
two benchmark suites demonstrate that while a typical binary will not experience
a slowdown of three orders of magnitude, microbenchmarks with small, tight loops
usually exhibit more significant slowdowns. This is convenient since the typical
application of performance degradation mechanisms is in conjunction with
side-channel attacks that target such hot spots.

In summary, the empirical data in this section validates the \HD concept and
answers \autoref{rq:hd} authoritatively. The data shows a clear advantage---even
reaching three orders of magnitude in select microbenchmark cases---of \HD
over classical \DG. Therefore, as a pure performance degradation mechanism, \HD
outperforms \DG.
\section{HyperDegrade: Assessment} \label{sec:eval}
Applying the \HD concept from \autoref{sec:concept}, \autoref{sec:perf}
subsequently showed the efficacy of \HD as a performance degradation technique.
Similar to the classical \DG technique, we see the main application of \HD in
the SCA area to improve the granularity of microarchitecture timing traces.
That is the focus of this section.

We first enumerate some of the shortcomings in previous work on performance
degradation. \citet[Sect.\ 5]{DBLP:conf/acsac/AllanBFPY16} show that decreasing
the \fl wait time---while indeed increasing granularity---generally leads to a
higher number of missed accesses concerning the targeted line. This was in fact
the main motivation for their \DG technique.
Applying \DG \cite[Sect.\ 7]{DBLP:conf/acsac/AllanBFPY16}, the authors argue why
missed accesses are detrimental to their end-to-end cryptanalytic attack. While
the intuition for their argument is logical, the authors provide no evidence,
empirical or otherwise, that \DG actually leads to traces containing
statistically more information, which is in fact the main purpose of performance
degradation techniques. The motivation and intuition by
\citet{DBLP:conf/uss/GarciaB17} is similar---albeit with a different framework
for target cache line identification---and equally lacks evidence.

The goal of this section is to answer \autoref{rq:nicv}, rectifying these shortcomings inspired by
information-theoretic methods. We do so by utilizing an established SCA metric
to demonstrate that classical \DG leads to statistically more leakage than \fl in
isolation. Additionally, our \HD technique further amplifies this leakage.

\subsection{Experiment}
\autoref{fig:victim} depicts the shared library we constructed to use throughout
the experiments in this section. The code has two functions \victimzero and
\victimone that are essentially the same, but separated by 512 bytes. The
functions set a counter (\code{r10}) from a constant (\code{CNT}, in this case
2k), then proceed through several effective nops (\code{add} and \code{sub}
instructions that cancel), then finally decrement the counter and iterate.

\begin{figure}
\scriptsize
\begin{alltt}
1200 <x64_victim_0>:               1400 <x64_victim_1>:
1200: mov $CNT,\RIO                1400: mov $CNT,\RIO
1207: add $0x1,\RIO                1407: add $0x1,\RIO
120b: sub $0x1,\RIO                140b: sub $0x1,\RIO
...                                ...
12e7: add $0x1,\RIO                14e7: add $0x1,\RIO
12eb: sub $0x1,\RIO                14eb: sub $0x1,\RIO
12ef: sub $0x1,\RIO                14ef: sub $0x1,\RIO
12f3: jnz 1207 <x64_victim_0+0x7>  14f3: jnz 1407 <x64_victim_1+0x7>
12f9: retq                         14f9: retq
\end{alltt}
\vspace{-4ex}
\caption{Functions of a shared library (\code{objdump} view) used to construct
an ideal victim for our SCA leakage assessment experiments.}
\label{fig:victim}
\end{figure}

We designed and implemented an ideal victim application linking against this
shared library. The victim either makes two sequential \victimzero calls
(``0-0'') or \victimzero followed by \victimone (``0-1''). We then used the
stock \fl technique, probing the start of \victimzero (i.e.\ at hex offset
\code{1200}).

Pinning the victim and spy to separate physical cores, we then procured 20k
traces, in two sets of 10k for each of 0-0 and 0-1, and took the mean of the
sets to arrive at the average trace. \autoref{fig:nicv} (Top) represents these
two baseline \fl cases with \NO strategy as the two plots
on the far left.

The next experiment was analogous, yet with the classical \DG strategy. We
degraded two cache lines---one in \victimzero and the \victimone, both in the
middle of their respective functions. These are the two middle plots in
\autoref{fig:nicv} (Top). Here the victim, spy, and degrade processes are all
pinned to different physical cores.

Our final experiment was analogous, yet with our novel \HD strategy and pinning
the victim and degrade processes to two logical cores of the same physical
core---degrading the same two cache lines---and the spy to a different physical core.
These are the two plots on the far right in \autoref{fig:nicv} (Top).

What can be appreciated in \autoref{fig:nicv} (Top), is that both performance
degradation strategies are working as intended---they are stretching the traces.
The remainder of this section focuses on quantifying this effect.
In fact \HD stretches the traces to such an extreme that the \NO data on the far
left is scantily discernible in this visualization.

\subsection{Results}
\begin{figure}[!t]
\includegraphics[width=\linewidth]{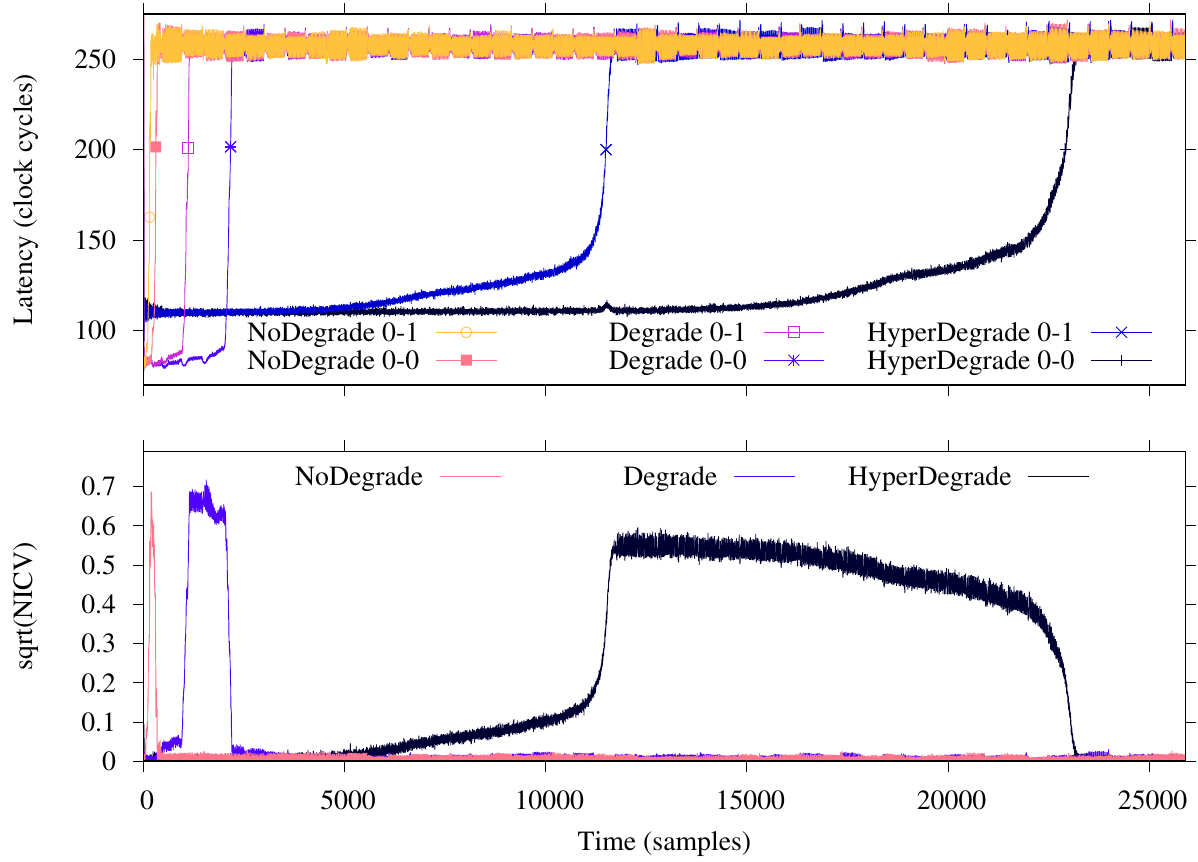}
\vspace{-4ex}
\caption{Top: averaged traces across different degrade strategies and different
victim execution paths (i.e.\ classes, 0-0 and 0-1). The legend corresponds to
the plots from left to right.
Bottom: the NICV metric's square root, or maximum correlation.
The legend again corresponds to the plots from left to right.
The plots align and display the same time slice.}
\label{fig:nicv}
\end{figure}

Recalling from \autoref{sec:leakage_assess}, NICV suits particularly well for
our purposes, since it is designed to work with only public data and is agnostic
to leakage models \cite{2014:NICV}. The latter fact makes NICV pertinent as a
metric to compare the quality of traces \cite[Sect.\ 3]{2014:NICV}.
The metric---in the interval $[0,1]$---is defined by
\begin{equation} \label{eq:nicvn}
\mathrm{NICV}(X,Y) = \frac{\mathrm{Var}[\mathrm{E}[Y | X]]}{\mathrm{Var}[Y]}
\end{equation}
with traces $Y$, classes $X$, and $\mathrm{E}$ the expectation (mean). The
square root of the NICV metric, or the correlation ratio, is an upper bound for
Pearson's correlation coefficient \cite[Corollary 8]{DBLP:journals/tc/ProuffRB09}.
Two classes (0-0 and 0-1) suffice for our purposes, simplifying
\autoref{eq:nicvn} as follows.
\begin{equation*} \label{eq:nicv2}
\mathrm{NICV}(X,Y) = \frac{(\mathrm{E}[Y | X=0] - \mathrm{E}[Y|X=1])^2}{4 \cdot \mathrm{Var}[Y]}
\end{equation*}
\autoref{fig:nicv} (Bottom) illustrates applying this metric to the two sets of
measurements for each degrade strategy---baseline \NO, \DG, and \HD---and
visualizing the square root, or maximum correlation. With simple thresholding to
identify POIs (i.e., those points that exceed a fixed value),
this leads to the POI statistics in \autoref{tab:pois}.
To give one extremely narrow interpretation, with \code{CNT} set to 2k
in \autoref{fig:victim}, less than 2k POIs indicates information is being lost,
i.e.\ the victim is running faster than the spy is capable of measuring. With
this particular victim in this particular environment, it implies neither \NO
nor \DG achieve sufficient trace granularity to avoid information loss, while
\HD does so with ease.

In conclusion, this definitively answers \autoref{rq:nicv}; the \DG
strategy leads to statistically more information leakage over stock \fl due to
the significant POI increase. Similarly, it shows
\HD leads to significantly more POIs compared to \DG, but at the same
time (on average) slightly lower maximum correlation for each POI.

\begin{table}[!t]
\caption{POI counts and ratios at various NICV thresholds across degrade
strategies (see \autoref{fig:nicv}).
The ratios (x) are between the different strategies.}
\label{tab:pois}
\centering
\resizebox{1.0\linewidth}{!}{%
\begin{tabular}{rrrr} \hline
Threshold & \NO & \DG & \HD \\ \hline
0.1 & 233 & 1212 (5.2x) & 13151 (56.4x, \BBBPAD{10.9x}) \\
0.2 & 188 & 1149 (6.1x) & 11664 (62.0x, \BBBPAD{10.2x}) \\
0.3 & 167 & 1097 (6.6x) & 11159 (66.8x, \BBBPAD{10.2x}) \\
0.4 & 147 & 1049 (7.1x) & 10194 (69.3x, \BBBPAD{9.7x}) \\
0.5 & 117 & 969 (8.3x) & 6003 (51.3x, \BBBPAD{6.2x}) \\
 \hline
\end{tabular}
}
\end{table}
\section{HyperDegrade: Exploitation} \label{sec:attack}
While \autoref{sec:eval} shows that \HD leads to more leakage due to the
significant increase in POIs, the \autoref{fig:victim} shared library and
linking victim application are unquestionably purely synthetic. While this is
ideal for leakage assessment, it does not represent the use of \HD in a real
end-to-end SCA attack scenario.
What remains is to demonstrate that \HD applies in end-to-end attack scenarios
and that \HD has a quantifiable advantage over other degrade strategies wrt.\
attacker effort. That is the purpose of this section.

\Paragraph{The leak}
Recalling \autoref{sec:raccoon}, the original Raccoon attack exploits the fact
that Diffie-Hellman as used in TLS 1.2 and below dictates stripping leading
zeros of the shared DH key during session key derivation. The authors note that
\emph{not} stripping is not foolproof can also lead to oracles \cite[Sect.\ 3.5]{tmp:raccoon},
pointing at an OpenSSL function that is potentially vulnerable to
microarchitecture attacks \cite[Appx.\ B]{tmp:raccoon}. They leave the
investigation of said function---unrelated to TLS---as future work: a gap which this section fills.

\begin{figure}
    \includegraphics[width=\linewidth]{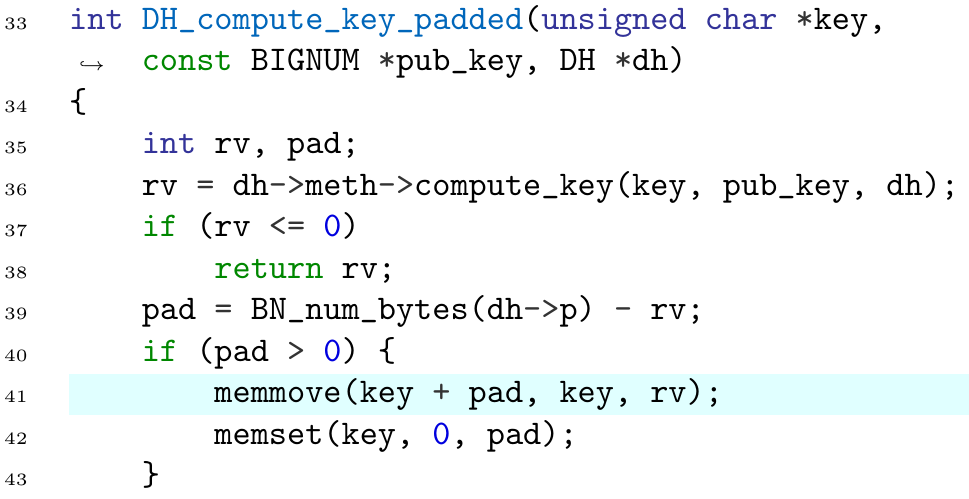}
    \caption{The target vulnerability in OpenSSL 1.1.1h Diffie-Hellman shared key derivation for our end-to-end attack.}
    \label{fig:dh_key}
\end{figure}

\autoref{fig:dh_key} shows that function, which is our target within the current (as of this writing)
state-of-the-art OpenSSL 1.1.1h DH shared secret key derivation. The shared
secret is computed at line 36; however, OpenSSL internals strip the leading zero
bytes of this result.
Therefore, at line 40 this function checks if the computed shared secret needs to be padded.
Padding is needed if the number of bytes of the shared secret and the DH modulus differ.

\Paragraph{The leakage model}
Considering a theoretical leakage model, the binary result of the line 40
condition leaks whether the shared secret has at least eight leading zero bits
(branch taken) or not (branch not taken).
This model---capable of extracting at most eight bits of information---affects
the key sizes that are in scope.
While 2048/256-bit DH parameters are more consistent with current key size
recommendations, the original Raccoon attack \cite[Table 3]{tmp:raccoon} is
unable to target eight bits of leakage in this setting: the authors explicitly
leave it as an open problem \cite[Sect.\ 6.2]{tmp:raccoon}. They instead target
legacy (1024/160-bit) or non-standard (1036/160-bit) DH parameters for an
eight-bit leak. We follow suit, targeting legacy keys
(see \autoref{sec:threatmodel}) for the exact same reasons.

\Paragraph{The victims}
Our next task was to identify callers to the \autoref{fig:dh_key}
code from the application and protocol levels, since it is unrelated to TLS.
We successfully identified PKCS~\#7
(\rfc{2315}) and CMS (\rfc{5652}) as standards where \autoref{fig:dh_key}
might apply. We subsequently used the TriggerFlow tool \cite{DBLP:conf/dimva/GridinGTB19}
to verify that OpenSSL's \code{cms} and \code{smime} command line utilities have
the \autoref{fig:dh_key} function in their call stacks.

\subsection{Attack Outline and Threat Model} \label{sec:threatmodel}

In our end-to-end attack, all message encryptions and decryptions are with
OpenSSL's command line \code{cms} utility. We furthermore assume Alice has a
static DH public key in an X.509 certificate and, wlog., the DH parameters are
the fixed 1024/160-bit variant from \rfc{5114}. OpenSSL supports these natively
as named parameters, used implicitly. We carried out all experiments on the
Coffee Lake machine from \autoref{tab:machines}.

Our Raccoon attack variant consists of the following steps.
(i) Obtain a target CMS-encrypted message from Bob to Alice.
(ii) Based on the target, construct many chosen ciphertexts and submit them to
Alice for decryption.
(iii) Monitor Alice's decryptions of these ciphertexts with \HD and \fl to
detect the key-dependent padding.
(iv) Use the resulting information to construct a lattice problem and recover
the original target session key between Bob and Alice, leading to loss of
confidentiality for the target message.
The original Raccoon attack \cite{tmp:raccoon} abstracts away most of these
steps, using only simulated SCA data.

\Paragraph{Threat model}
Our attack makes several assumptions discussed below, which we borrow directly
from the existing literature.
(i) Our threat model assumes the attacker is able to co-locate on the same system
with Alice (victim), and furthermore execute on the same logical and physical
cores in parallel to Alice. See the end of \autoref{sec:concept} for a
discussion of this standard assumption.
(ii) We also assume that Alice decrypts messages non-interactively, due to the number
of queries required. This is a fair assumption not only because DH is literally
Non-Interactive Key Exchange (NIKE) \cite{DBLP:conf/pkc/FreireHKP13} from the
theory perspective, but also because CMS (the evolution of PKCS~\#7) has ubiquitous use cases,
e.g.\ including S/MIME.
Chosen ciphertext decryptions is a standard assumption from the applied SCA literature
\cite[Sect.\ 1.1]{DBLP:conf/crypto/GenkinST14}
\cite[Sect.\ 1.4]{DBLP:conf/ccs/GenkinVY17}
\cite[Sect.\ 3]{DBLP:conf/sp/RonenGGSWY19}.
(iii) We assume the attacker is able to observe one encrypted message from Bob
to Alice. This is a passive variant of the standard Dolev-Yao
adversary \cite{DBLP:journals/tit/DolevY83} that is Man-in-the-Middle (MitM)
capable of eavesdropping, and the exact same assumption from the original
Raccoon attack \cite[Fig.\ 1]{tmp:raccoon}.
\citet[Sect.\ 3]{DBLP:conf/sp/RonenGGSWY19} call this
\emph{privileged network position} since it is a weak assumption compared to
full MitM capabilities.
To summarize, the overall threat model used by \citet[Sect.\ 3]{DBLP:conf/sp/RonenGGSWY19}
is extremely similar to ours and encompasses all of the above assumptions. The
only slight difference is a stronger notion of co-location in our case---from
same CPU to same physical core.

\Paragraph{Case study: triggering oracle decryptions}
We briefly explored the non-interactive requirement discussed above.
Specifically, two arenas: automated email decryption, and automated decryption
of certificate-related messages.

Recent changes in Thunderbird (v78+) migrate from the Enigmail plugin to native
support for email encryption and/or authentication (PGP, S/MIME). Automated,
non-interactive decryption for various purposes (e.g., filtering) appears to be
a non-default (yet supported) option.%
\footurl{https://bugzilla.mozilla.org/show_bug.cgi?id=1644085}
Quoting from that thread: ``A lot of companies e.g.\ in the finance sector
decrypt the messages at a central gateway and then forward them internally to
the respective recipient.''

We also found explicit code meeting our non-interactive requirement in the realm
of automated certificate management.
(i) The Simple Certificate Enrollment Protocol
(SCEP, \rfc{8894}) supports exchanging (public key encrypted, CMS formatted)
confidential messages over an insecure channel, such as HTTP or generally
out-of-band. This is in contrast to the Automatic Certificate Management
Environment (ACME) protocol (\rfc{8555}, e.g.,
Let's Encrypt \cite{DBLP:conf/ccs/AasBCDEFHHKRSW19}), which relies on the
confidentiality and authenticity guarantees of TLS. The open
source\footurl{https://redwax.eu/rs/docs/latest/mod/mod_scep.html}
Apache module \code{mod\_scep} dynamically links against OpenSSL to provide this
functionality.
(ii) The Certificate Management Protocol (CMP, \rfc{4210}) provides
similar functionality (i.e., public key encrypted, CMS formatted messages) with
similar motivations (automated certificate management over insecure channels).
Yet the implementation integrated into upcoming OpenSSL 3.0 does not currently
support encrypted protocol messages.%
\footurl{https://www.openssl.org/docs/manmaster/man1/openssl-cmp.html}

\subsection{Degrade Strategies Compared} \label{sec:cmp}
This section aims at answering \autoref{rq:realHD}
by means of comparing three performance degradation strategies (\NO, \DG, \HD)
when paired with a \fl attack to exploit this vulnerability.
We reuse the following setup and adversary plan later during the end-to-end attack (\autoref{sec:attack2}).

\Paragraph{Experiment}
We monitor the cache line corresponding to the \code{memmove} function call and its
surrounding instructions, i.e.\ near line 41 of \autoref{fig:dh_key}.
If \code{memmove} is executed, at least two cache hits should be observed:
(i) when the function is called,
(ii) then when the function finishes (\code{ret} instruction).
Therefore, if two cache hits are observed in a trace \emph{close} to each other,
that would mean the shared secret was padded, and in contrast a single cache hit
only detects flow surrounding line 41 of \autoref{fig:dh_key}.

We select the first cache line where the function \code{memmove} is located as the degrading cache line.
It is the stock, unmodified, uninstrumented \code{memmove}
available system-wide as part of the shared C standard library \code{libc}.
Degrading during \code{memmove} execution should increase the time window
the spy process has to detect the second cache hit (\ie increase time granularity).

We strive for a \emph{fair} comparison between the three degradation strategies
during a \fl attack.
It is challenging to develop an optimal attack for each degradation strategy and even harder to maintain fairness.
Therefore, we developed a single attack plan and swept its parameters in order to provide a meaningful and objective comparison.

\autoref{tab:params} summarizes the attack parameters and the explored search space.
The first parameter, $r$, affects trace capturing---it
specifies the number of iterations the \fl wait loop should iterate.
The remaining parameters belong to the trace processing tooling.
The second parameter, $t$, refers to the threshold in CPU clock cycles used to distinguish a cache hit from a miss.
After some manual trace inspection, we observed this threshold varies
between degradation strategy and \fl wait time; we decided to add it to the search space.
The last parameter, $d$, specifies the distance (in number of \fl samples) between two cache hits to
consider them as close.

\begin{table}
    \caption{Attack parameters search space.}
    \label{tab:params}
    \centering
    \begin{tabular}{ll} \hline
        Parameter & Range         \\ \hline
        \fl wait time ($r$) & $\{128, 256\}$        \\
        Cache hit/miss threshold ($t$) & $\{50, 100, 150, 200\}$ \\
        Cache hits closeness distance ($d$) & $\{1, 5, 10, \ldots, 95\}$      \\ \hline
    \end{tabular}
\end{table}

We explore this parameter search space, and for each parameter set---\ie triplet $(r, t, d)$---evaluate
the attack performance, estimating the true positive (TP) and false positive rates (FP).
For this task, we generated two pairs of DH keys (\ie attacker and victim).
We selected one of these pairs such that the shared secret needs padding after a DH key exchange,
while for the other it does not.

We then captured 1k traces for each key pair, parameter set, and degradation strategy under consideration
and estimated the TP and FP rates.
We are interested in finding which parameter sets lead to more efficient attacks in terms of number of
traces to capture, i.e.\ number of attacker queries.
Therefore, we focused on those results with zero false positives for the comparison,
thus it is a best case analysis for all degradation strategies.

\Paragraph{Results}
For 1024-bit DH, the lattice-based cryptanalysis requires 173 samples where padding occurred (explained later).
Therefore, the following equation defines the average number of traces that need to be captured,
where $\Pr[\text{pad}] = 1/177 \approx 0.00565$ with the fixed \rfc{5114} parameters.
\begin{equation*}\label{eq:num_traces}
     \text{num\_traces} = 173 / (\Pr[\text{TP}] \cdot \Pr[\text{pad}])
\end{equation*}
Considering the very low probability the leakage occurs
and the increased complexity of lattice-based cryptanalysis in the
presence of errors (see \cite[Sect.\ 6]{DBLP:conf/eurocrypt/AlbrechtH21}),
reducing the number of traces is important for attack effectiveness.

\begin{figure}
\includegraphics[width=\linewidth]{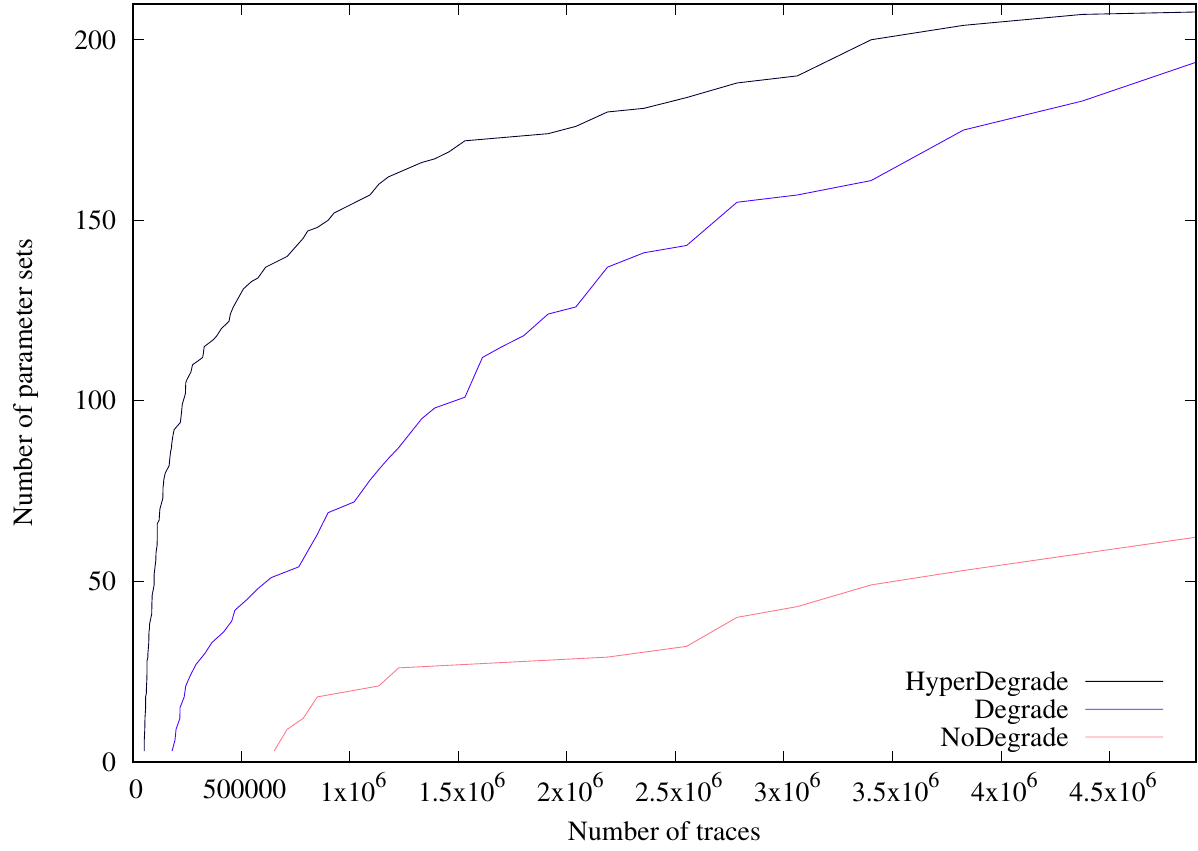}
\vspace{-4ex}
\caption{Degrade strategies comparison: how many parameter sets can be used to mount an attack using $x$-number of traces.}
\label{fig:hist_cmp}
\end{figure}

\autoref{tab:bests} shows the best parameter set results for each degrade strategy
that could lead to a successful attack.
Note that \HD clearly reduced the number of required traces to succeed by at least a factor of $3.3$
when compared with \DG (the second best performer).
This translates into a considerable reduction in the number of traces: from 181k to 53k.
Moreover, \autoref{fig:hist_cmp} shows there is not just a single parameter set
where \HD performs better than \DG, but rather there are $88$ of them.
These results provide evidence that \HD can perform better than the other
two degrade strategies for mounting \fl attacks on cryptography applications,
answering \autoref{rq:realHD}.

\begin{table}
    \caption{Best results for degrade strategies.}
    \label{tab:bests}
    \centering
    \begin{tabular}{lcl} \hline
        Strategy & Trace count & Param.\ set $(r,d,t)$ \\ \hline
        \NO &      651510  &  $(128, 1, 100)$  \\
        \DG &      181189  &  $(256, 1, 170)$  \\
        \HD & \,\,\,53721  &  $(256, 1, 170)$  \\ \hline
    \end{tabular}
\end{table}

\subsection{End-to-End Attack Instance} \label{sec:attack2}

The remainder of this section answers \autoref{rq:raccoon}.
We begin with lattice details, then finish with the results of our end-to-end attack.

\Paragraph{Lattice construction}
Alice's public
key $g^a$ is readily available and the attacker observes $g^b$ from the original
target query, along with ciphertext encrypted under the shared session key
$g^{ab}$ (private). Then the attacker proceeds with chosen queries, crafting
ciphertext $g^b g^{r_i}$ and random $r_i$ for submitting to Alice for
decryption. Alice then computes $(g^b g^{r_i})^a = g^{ab} \cdot (g^a)^{r_i}$
with the attacker measuring if padding occurs.
This is an instance of the hidden number problem (HNP) by \citet{DBLP:conf/crypto/BonehV96}---to
recover $\alpha = g^{ab}$ given many $t_i = (g^a)^{r_i}$.

We use the lattice construction by \citet{DBLP:journals/dcc/NguyenS03} verbatim,
stated here for completeness.
Restricting to the $t_i$ where padding occurred,
our SCA data tells us $0 < \alpha t_i < p/{2^{\ell}}$, where
we set $\ell=8$ due to the nature of this particular side channel; recall
``branch taken'' in \autoref{fig:dh_key} says at least the top eight bits are clear.
Denoting $u_i = p/2^{\ell + 1}$ yields
$v_i = \lvert \alpha t_i - u_i \rvert_p \leq p/{2^{\ell+1}}$
where $\lvert x \rvert_p$ is signed modulo $p$
reduction centered around zero. Then there are integers $\lambda_i$ where
$\abs(\alpha t_i - u_i - \lambda_i) \leq p/{2^{\ell+1}}$ holds, and this is the
key observation for lattice attacks; the $u_i$ approximate $\alpha t_i$ since
they are closer than a random integer modulo $p$.
Consider the rational $d+1$-dimension lattice generated by the rows of the
following matrix.
\[
B =
\begin{bmatrix}
2 W p & 0 & \dots & \dots & 0 \\
0 & 2 W p & \ddots & \vdots & \vdots \\
\vdots & \ddots & \ddots & 0 & \vdots \\
0 & \dots & 0 & 2 W p & 0 \\
2 W t_1 & \dots & \dots & 2 W t_d & 1
\end{bmatrix}
\]
When we set $W = 2^\ell$, $\vec{x}=(\lambda_1,\ldots,\lambda_d,\alpha)$,
$\vec{y}=(2 W v_1,\ldots,2 W v_d,\alpha)$, and
$\vec{u}=(2 W u_1,\ldots,2 W u_d,0)$ we get the linear relationship
$\vec{x}B-\vec{u}=\vec{y}$.
Solving the Closest Vector Problem (CVP) with inputs $B$ and $\vec{u}$ yields
$\vec{x}$, and hence the target session key $\alpha$. We also use the traditional
CVP-to-SVP (Shortest Vector Problem) embedding by \citet[Sec.\ 3.4]{DBLP:conf/crypto/GoldreichGH97a}.
\citet{DBLP:conf/uss/GarciaHTGAB20} suggest weighting on
the average logarithm, hence we set $W=2^{\ell + 1}$ in our $\ell=8$ scenario.
Lastly, to set the lattice dimension
we use the heuristic from \cite[Sect.\ 9.1]{DBLP:conf/ccs/HassanGDGCAB20} verbatim,
which is $d = 1-e^{-c}$. With their suggested confidence factor $c=1.35$ (to
improve key bit independence), in our case it leads to $d=173$; this explains
the constant from \autoref{sec:cmp}.
We use the same BKZ block size parameter as \cite[Table 3]{tmp:raccoon}, $\beta=60$.

\Paragraph{Results}
Following the results of \autoref{sec:cmp}, we proceeded to capture 60k traces using \HD
and the parameter set shown in \autoref{tab:bests}. Our capture tooling
implements precisely step (i) to (iii) in \autoref{sec:threatmodel}, obtaining a
target ciphertext, constructing chosen ciphertexts, and querying the oracle. In
the end, at each capture iteration the attacker only needs to modify a public
key field in an ASN.1 structure to produce a new chosen ciphertext, then take
the measurements while Alice performs the decryption.

Considering the padding probability $\Pr[\text{pad}] = 1/177$,
the expected number of padded traces in the $60\text{k}$ set is $339$.
After processing each trace, our tooling detected $3611$ paddings.
It indicates that, with high probability, these also contain false positives.
\citet{DBLP:conf/eurocrypt/AlbrechtH21} focus on lattice-based cryptanalysis of
ECDSA and suggest adjusting lattice parameters at the cost of increased
computation to compensate for errors. We instead use a different approach in the
DH setting---not applicable in the ECDSA setting due to its usage of nonces---to
counteract those false positives.

To reduce the FP rate, for each trace where we detected padding,
we retry the query seven times and majority vote the result.
From the $3611$ traces detected as padded, only $239$ passed the majority voting.
Therefore, the total number of traces captured was $60\text{k} + 7 \cdot 3611 = 85277$.

Even with the majority voting, some false positives could remain;
we sorted the $239$ samples by the vote count.
Then we selected the highest ranked 173 samples to build the HNP instances.
For the sake of completeness, we verified there were 47 false positives in the $239$ set;
however, all had the lowest vote count of four.

We implemented our lattice using BKZ reduction from
\code{fpylll}\footurl{https://github.com/fplll/fpylll}, a Python wrapper for the
\code{fplll} C++ library \cite{fplll}. We constructed 24 lattice instances from
our SCA data, and executed these in parallel on a 2.1 GHz dual CPU Intel Xeon
Silver 4116 (24 cores, 48 threads across 2 CPUs) running Ubuntu 20 with 256 GB
memory. The first instance to recover the session key did so in one hour and
five minutes with a single BKZ reduction. With no abstractions, utilizing real
trace data at the application level, and real protocol messages, our end-to-end
attack resolves \autoref{rq:raccoon}.

\Paragraph{OpenSSL disclosure}
We contacted the OpenSSL security team to disclose our results regarding the
exploitability of this leak.
We also designed, implemented and tested a fix\footurl{https://github.com/openssl/openssl/pull/13772}
that avoids executing the leaky branch.
We achieved this by changing the default behavior of the \code{dh->meth->compute\_key}
function pointer to always return a fixed-length array
(i.e., the public byte length of $p$) in constant time,
ensuring variable \code{pad} is zero in \autoref{fig:dh_key}
(i.e., \code{BN\_num\_bytes(dh->p)} and \code{rv} are equal).
Retaining the branch and not simply removing it is due to backwards compatibility
issues with OpenSSL engines \cite{DBLP:conf/secdev/TuveriB19}.
OpenSSL merged our fix on 10 January 2021, included as of version 1.1.1j.
\section{Conclusion} \label{sec:conclusion}

\HD increases performance degradation with respect to state-of-the-art.
The difference depends on the targeted process, but we achieved slowdown factors
up to three orders of magnitude in select microbenchmark cases.
In addition to increased cache misses, we discovered the cache-based performance
degradation root cause is due to the increased number of machine clears produced
by the processor detecting a cache line flush from L1 as self-modifying code
(\autoref{rq:hd}).
We analyzed the impact of \DG and \HD on \fl traces from a theoretical point of view
using leakage assessment tools, demonstrating that \HD tremendously increases
the number of POIs, which reflects in an increased time granularity
(\autoref{rq:nicv}).
From an applied perspective, we designed a fair experiment that compares the three
degrade strategies \NO, \DG, and \HD when coupled with a \fl attack wrt.\ the
number of traces needed to recover a secret from a cryptography implementation
(\autoref{rq:realHD}). Our resulting data demonstrates the benefits of \HD,
requiring three times less traces and attacker queries to succeed, the latter
being the standard metric in applied SCA literature.
Regarding cryptography, we answered an open problem from the recently published
Raccoon attack, providing experimental evidence that such an attack applies
with real data (\autoref{rq:raccoon}).

\Paragraph{Future work}
Our work either reinforces or illuminates several new avenues for continued related research.

In \autoref{sec:cmp}, we noted how the cache hit threshold varies depending on various spy parameters.
We have also noted this behavior in other \fl scenarios, outside this work.
It would definitely be an interesting future research line to investigate its root cause.

In general, our off-the-shelf applied lattice techniques in \autoref{sec:attack2},
while serving their purpose for proof-of-concept, are likely not optimal.
Fundamental lattice-based cryptanalysis improvements
(e.g., the recent \cite{DBLP:conf/eurocrypt/AlbrechtH21}) are beyond the scope of our work,
but could reduce dimension and subsequently attacker queries.
Similar to our Raccoon variant, the original Raccoon attack
\cite{tmp:raccoon} is unable to target 2048/256-bit DH with eight bits or less
of leakage. The authors leave this as an open problem, and we concur; indeed,
improved lattice methods to compensate for these significantly larger finite
field elements is an interesting research direction.

Several previous studies gather widespread certificate and key usage statistics.
For example, the original Raccoon attack authors gather statistics for static DH keys
in X.509 certificates for TLS 1.2 (and lower) authentication, and/or
ephemeral-static DH keys in TLS 1.2 (and lower) cipher
suites \cite[Sect.\ 7]{tmp:raccoon} from public services.
\citet{DBLP:conf/fc/BosHHMNW14} gather publicly-available elliptic curve keys
from protocols and services such as TLS, SSH, BitCoin, and the Austrian
e-identity card; \citet{DBLP:conf/eurosp/ValentaSSH18} consider IPSec, as well.
Not specific to any particular public key cryptosystem,
\citet{DBLP:conf/crypto/LenstraHABKW12} gather publicly-available PGP keys and
X.509 certificates for TLS 1.2 (and lower) authentication.
Along these lines, although it is beyond the scope of our work, we call for
future studies that gather and share S/MIME key usage statistics, paying
particular attention to legal issues and privacy since, different from PGP,
we are not aware of any general public (distributed) repositories for S/MIME keys.

\Paragraph{Acknowledgments}
This project has received funding from the European Research Council (ERC) under
the European Union's Horizon 2020 research and innovation programme (grant
agreement No 804476).
Supported in part by CSIC's i-LINK+ 2019 ``Advancing in cybersecurity technologies''
(Ref.\ LINKA20216).
 
\begin{table*}
\caption{BEEBS performance degradation results (cycles, thousands) on Skylake and Kaby Lake.}
\label{tab:beebs2}
\centering
\resizebox{0.85\textwidth}{!}{%
\begin{tabular}{|l|rrrrr|rrrrr|} \hline
& \multicolumn{5}{|c|}{Skylake} & \multicolumn{5}{|c|}{Kaby Lake} \\
Benchmark &
\NO &
\multicolumn{2}{c}{\DG} &
\multicolumn{2}{c|}{\HD} &
\NO &
\multicolumn{2}{c}{\DG} &
\multicolumn{2}{c|}{\HD} \\ \hline
\texttt{aha-compress} &
70583 &
848723 & (12.0x) &
16222580 & (229.8x) &
69754 &
697353 & (10.0x) &
18585074 & (266.4x) \\
\texttt{aha-mont64} &
21954 &
463471 & (21.1x) &
12066723 & (549.6x) &
22148 &
356097 & (16.1x) &
11607976 & (524.1x) \\
\texttt{bs} &
2032 &
8991 & (4.4x) &
171277 & (84.3x) &
2180 &
7137 & (3.3x) &
152106 & (69.8x) \\
\texttt{bubblesort} &
332386 &
9244976 & (27.8x) &
249121317 & (749.5x) &
305248 &
9197521 & (30.1x) &
229279506 & (751.1x) \\
\texttt{cnt} &
13237 &
191420 & (14.5x) &
4482733 & (338.6x) &
13488 &
187061 & (13.9x) &
5202446 & (385.7x) \\
\texttt{compress} &
8878 &
104299 & (11.7x) &
3248479 & (365.9x) &
9001 &
109617 & (12.2x) &
3767946 & (418.6x) \\
\texttt{cover} &
6982 &
172070 & (24.6x) &
2777996 & (397.9x) &
7165 &
95582 & (13.3x) &
2592937 & (361.9x) \\
\texttt{crc} &
7671 &
139690 & (18.2x) &
3745933 & (488.3x) &
7839 &
152098 & (19.4x) &
3516855 & (448.6x) \\
\texttt{crc32} &
46875 &
1458728 & (31.1x) &
31199559 & (665.6x) &
47004 &
1716899 & (36.5x) &
33339067 & (709.3x) \\
\texttt{ctl-stack} &
37481 &
528762 & (14.1x) &
11150759 & (297.5x) &
38447 &
577541 & (15.0x) &
15785236 & (410.6x) \\
\texttt{ctl-string} &
31088 &
981144 & (31.6x) &
14581397 & (469.0x) &
32189 &
799759 & (24.8x) &
17316754 & (538.0x) \\
\texttt{ctl-vector} &
30742 &
367393 & (12.0x) &
7275579 & (236.7x) &
31266 &
372204 & (11.9x) &
8280240 & (264.8x) \\
\texttt{cubic} &
33333 &
201498 & (6.0x) &
3240751 & (97.2x) &
30686 &
184586 & (6.0x) &
4497482 & (146.6x) \\
\texttt{dijkstra} &
1965916 &
39394667 & (20.0x) &
962126944 & (489.4x) &
1979452 &
36038128 & (18.2x) &
1032841427 & (521.8x) \\
\texttt{dtoa} &
13236 &
76977 & (5.8x) &
1374173 & (103.8x) &
13422 &
77959 & (5.8x) &
1838910 & (137.0x) \\
\texttt{duff} &
6332 &
60486 & (9.6x) &
2488543 & (393.0x) &
6448 &
56663 & (8.8x) &
1876609 & (291.0x) \\
\texttt{edn} &
192602 &
4037466 & (21.0x) &
65758984 & (341.4x) &
191705 &
2978126 & (15.5x) &
86795945 & (452.8x) \\
\texttt{expint} &
29724 &
395738 & (13.3x) &
4513790 & (151.9x) &
30193 &
377277 & (12.5x) &
5138636 & (170.2x) \\
\texttt{fac} &
4595 &
64388 & (14.0x) &
1719178 & (374.1x) &
4761 &
60222 & (12.6x) &
1815137 & (381.2x) \\
\texttt{fasta} &
2218271 &
28380647 & (12.8x) &
628593367 & (283.4x) &
2216456 &
33215414 & (15.0x) &
644852892 & (290.9x) \\
\texttt{fdct} &
7759 &
23118 & (3.0x) &
935889 & (120.6x) &
7863 &
33953 & (4.3x) &
993673 & (126.4x) \\
\texttt{fibcall} &
2889 &
19083 & (6.6x) &
751337 & (260.0x) &
3054 &
19906 & (6.5x) &
824186 & (269.8x) \\
\texttt{fir} &
764841 &
18665980 & (24.4x) &
731941919 & (957.0x) &
764893 &
17441064 & (22.8x) &
685505153 & (896.2x) \\
\texttt{frac} &
12426 &
138238 & (11.1x) &
3326194 & (267.7x) &
12614 &
153789 & (12.2x) &
3927731 & (311.4x) \\
\texttt{huffbench} &
1400373 &
12636520 & (9.0x) &
185693684 & (132.6x) &
1424397 &
10053959 & (7.1x) &
187339402 & (131.5x) \\
\texttt{insertsort} &
4381 &
76538 & (17.5x) &
1937214 & (442.1x) &
4518 &
82371 & (18.2x) &
2423159 & (536.3x) \\
\texttt{janne\_complex} &
2443 &
17672 & (7.2x) &
387808 & (158.7x) &
2589 &
15907 & (6.1x) &
403711 & (155.9x) \\
\texttt{jfdctint} &
11742 &
117469 & (10.0x) &
2469100 & (210.3x) &
11917 &
106261 & (8.9x) &
3063148 & (257.0x) \\
\texttt{lcdnum} &
2247 &
20278 & (9.0x) &
283506 & (126.1x) &
2419 &
14790 & (6.1x) &
357861 & (147.9x) \\
\texttt{levenshtein} &
151336 &
3413532 & (22.6x) &
94467885 & (624.2x) &
148115 &
3324556 & (22.4x) &
92479156 & (624.4x) \\
\texttt{ludcmp} &
8941 &
86599 & (9.7x) &
2064081 & (230.8x) &
9190 &
71764 & (7.8x) &
2182355 & (237.5x) \\
\texttt{matmult-float} &
67852 &
1760965 & (26.0x) &
45235894 & (666.7x) &
68377 &
1481444 & (21.7x) &
53988799 & (789.6x) \\
\texttt{matmult-int} &
438567 &
13438448 & (30.6x) &
371621846 & (847.4x) &
444120 &
9045445 & (20.4x) &
424979930 & (956.9x) \\
\texttt{mergesort} &
519862 &
9598330 & (18.5x) &
179589127 & (345.5x) &
517294 &
8497198 & (16.4x) &
207490953 & (401.1x) \\
\texttt{miniz} &
3405 &
17987 & (5.3x) &
224863 & (66.0x) &
3547 &
12738 & (3.6x) &
290558 & (81.9x) \\
\texttt{minver} &
6391 &
53104 & (8.3x) &
1275079 & (199.5x) &
6824 &
44293 & (6.5x) &
1331331 & (195.1x) \\
\texttt{nbody} &
250992 &
5342357 & (21.3x) &
164185274 & (654.1x) &
253584 &
5439927 & (21.5x) &
162955284 & (642.6x) \\
\texttt{ndes} &
113555 &
1740050 & (15.3x) &
34799379 & (306.5x) &
119918 &
1563197 & (13.0x) &
53176247 & (443.4x) \\
\texttt{nettle-aes} &
113306 &
489386 & (4.3x) &
14676852 & (129.5x) &
113123 &
488887 & (4.3x) &
21739421 & (192.2x) \\
\texttt{nettle-arcfour} &
87327 &
1784265 & (20.4x) &
22378097 & (256.3x) &
87136 &
1515589 & (17.4x) &
25846101 & (296.6x) \\
\texttt{nettle-cast128} &
13957 &
23609 & (1.7x) &
198077 & (14.2x) &
13263 &
23692 & (1.8x) &
166519 & (12.6x) \\
\texttt{nettle-des} &
9179 &
27907 & (3.0x) &
250080 & (27.2x) &
9336 &
30056 & (3.2x) &
202710 & (21.7x) \\
\texttt{nettle-md5} &
7482 &
35234 & (4.7x) &
549028 & (73.4x) &
7604 &
29737 & (3.9x) &
732806 & (96.4x) \\
\texttt{nettle-sha256} &
14957 &
55160 & (3.7x) &
1459782 & (97.6x) &
15014 &
56483 & (3.8x) &
1314044 & (87.5x) \\
\texttt{newlib-exp} &
3667 &
23027 & (6.3x) &
344297 & (93.9x) &
3815 &
23708 & (6.2x) &
484350 & (126.9x) \\
\texttt{newlib-log} &
3176 &
14916 & (4.7x) &
352798 & (111.1x) &
3304 &
15313 & (4.6x) &
417836 & (126.4x) \\
\texttt{newlib-mod} &
2280 &
10949 & (4.8x) &
203870 & (89.4x) &
2486 &
11338 & (4.6x) &
221979 & (89.3x) \\
\texttt{newlib-sqrt} &
9448 &
140972 & (14.9x) &
5497784 & (581.8x) &
9608 &
143752 & (15.0x) &
3787532 & (394.2x) \\
\texttt{ns} &
27810 &
920498 & (33.1x) &
30642549 & (1101.9x) &
24270 &
718504 & (29.6x) &
25730490 & (1060.1x) \\
\texttt{nsichneu} &
12465 &
17453 & (1.4x) &
129175 & (10.4x) &
12680 &
17799 & (1.4x) &
128992 & (10.2x) \\
\texttt{prime} &
58915 &
934324 & (15.9x) &
11517642 & (195.5x) &
57920 &
922405 & (15.9x) &
14618565 & (252.4x) \\
\texttt{qsort} &
3868 &
34437 & (8.9x) &
636532 & (164.5x) &
4014 &
26922 & (6.7x) &
831894 & (207.2x) \\
\texttt{qurt} &
5456 &
56121 & (10.3x) &
972214 & (178.2x) &
5588 &
59091 & (10.6x) &
1089667 & (195.0x) \\
\texttt{recursion} &
7983 &
149886 & (18.8x) &
4806422 & (602.0x) &
7365 &
140916 & (19.1x) &
5193849 & (705.1x) \\
\texttt{rijndael} &
1831062 &
11743208 & (6.4x) &
235853943 & (128.8x) &
1830953 &
9728816 & (5.3x) &
280555490 & (153.2x) \\
\texttt{select} &
2499 &
18750 & (7.5x) &
286758 & (114.7x) &
2615 &
15980 & (6.1x) &
310185 & (118.6x) \\
\texttt{sglib-arraybinsearch} &
32695 &
941448 & (28.8x) &
24948746 & (763.1x) &
32073 &
913077 & (28.5x) &
25650916 & (799.7x) \\
\texttt{sglib-arrayheapsort} &
73813 &
964227 & (13.1x) &
28155013 & (381.4x) &
74316 &
1056981 & (14.2x) &
27101576 & (364.7x) \\
\texttt{sglib-arrayquicksort} &
35820 &
584648 & (16.3x) &
21360149 & (596.3x) &
35771 &
445614 & (12.5x) &
19761832 & (552.4x) \\
\texttt{sglib-dllist} &
103228 &
931802 & (9.0x) &
19577829 & (189.7x) &
103490 &
784713 & (7.6x) &
21655708 & (209.3x) \\
\texttt{sglib-hashtable} &
73515 &
720144 & (9.8x) &
13448462 & (182.9x) &
75750 &
672251 & (8.9x) &
16372567 & (216.1x) \\
\texttt{sglib-listinsertsort} &
148962 &
4155691 & (27.9x) &
57359384 & (385.1x) &
147850 &
4116964 & (27.8x) &
82759796 & (559.8x) \\
\texttt{sglib-listsort} &
82649 &
851066 & (10.3x) &
21974191 & (265.9x) &
83363 &
786974 & (9.4x) &
22087737 & (265.0x) \\
\texttt{sglib-queue} &
79274 &
976908 & (12.3x) &
20131755 & (254.0x) &
79976 &
999905 & (12.5x) &
28994719 & (362.5x) \\
\texttt{sglib-rbtree} &
194861 &
1795057 & (9.2x) &
39956917 & (205.1x) &
198968 &
1518989 & (7.6x) &
47673755 & (239.6x) \\
\texttt{slre} &
81356 &
961313 & (11.8x) &
22990277 & (282.6x) &
81068 &
885753 & (10.9x) &
27643530 & (341.0x) \\
\texttt{sqrt} &
278604 &
5427855 & (19.5x) &
66636453 & (239.2x) &
275896 &
4302655 & (15.6x) &
78476306 & (284.4x) \\
\texttt{st} &
46255 &
758154 & (16.4x) &
13044156 & (282.0x) &
45667 &
563498 & (12.3x) &
14099803 & (308.7x) \\
\texttt{statemate} &
5587 &
36137 & (6.5x) &
989900 & (177.2x) &
5768 &
36597 & (6.3x) &
1074219 & (186.2x) \\
\texttt{stb\_perlin} &
170355 &
2261082 & (13.3x) &
64860091 & (380.7x) &
141765 &
2404186 & (17.0x) &
63274557 & (446.3x) \\
\texttt{stringsearch1} &
15819 &
91324 & (5.8x) &
4618112 & (291.9x) &
16061 &
102834 & (6.4x) &
4935053 & (307.3x) \\
\texttt{strstr} &
5200 &
48645 & (9.4x) &
1224891 & (235.5x) &
5354 &
51939 & (9.7x) &
1345225 & (251.2x) \\
\texttt{tarai} &
2884 &
28585 & (9.9x) &
754962 & (261.7x) &
2965 &
18523 & (6.2x) &
682388 & (230.1x) \\
\texttt{trio-snprintf} &
16952 &
78512 & (4.6x) &
1100793 & (64.9x) &
17144 &
72171 & (4.2x) &
1273947 & (74.3x) \\
\texttt{trio-sscanf} &
21499 &
152353 & (7.1x) &
2933134 & (136.4x) &
22001 &
120026 & (5.5x) &
3577032 & (162.6x) \\
\texttt{ud} &
11215 &
69845 & (6.2x) &
1724508 & (153.8x) &
11106 &
75100 & (6.8x) &
2196806 & (197.8x) \\
\texttt{whetstone} &
335501 &
2891946 & (8.6x) &
55686230 & (166.0x) &
301591 &
2700492 & (9.0x) &
61531596 & (204.0x) \\
 \hline
\end{tabular}
}
\end{table*}

\begin{table*}
\caption{BEEBS performance degradation results (cycles, thousands) on Coffee Lake and Whiskey Lake.}
\label{tab:beebs3}
\centering
\resizebox{0.85\textwidth}{!}{%
\begin{tabular}{|l|rrrrr|rrrrr|} \hline
& \multicolumn{5}{|c|}{Coffee Lake} & \multicolumn{5}{|c|}{Whiskey Lake} \\
Benchmark &
\NO &
\multicolumn{2}{c}{\DG} &
\multicolumn{2}{c|}{\HD} &
\NO &
\multicolumn{2}{c}{\DG} &
\multicolumn{2}{c|}{\HD} \\ \hline
\texttt{aha-compress} &
71096 &
813585 & (11.4x) &
21102464 & (296.8x) &
69668 &
816475 & (11.7x) &
23975673 & (344.1x) \\
\texttt{aha-mont64} &
21931 &
468459 & (21.4x) &
12821078 & (584.6x) &
21687 &
492008 & (22.7x) &
15890278 & (732.7x) \\
\texttt{bs} &
1874 &
9676 & (5.2x) &
183221 & (97.8x) &
1763 &
8435 & (4.8x) &
207217 & (117.5x) \\
\texttt{bubblesort} &
335556 &
9468033 & (28.2x) &
322437859 & (960.9x) &
302489 &
11286756 & (37.3x) &
369824760 & (1222.6x) \\
\texttt{cnt} &
13046 &
231693 & (17.8x) &
5177506 & (396.8x) &
13030 &
240288 & (18.4x) &
6270126 & (481.2x) \\
\texttt{compress} &
8728 &
115043 & (13.2x) &
4971456 & (569.6x) &
8552 &
115566 & (13.5x) &
5197671 & (607.7x) \\
\texttt{cover} &
6779 &
115233 & (17.0x) &
3332393 & (491.6x) &
7151 &
118538 & (16.6x) &
2716975 & (379.9x) \\
\texttt{crc} &
7526 &
160400 & (21.3x) &
4230637 & (562.1x) &
7601 &
177394 & (23.3x) &
3880102 & (510.4x) \\
\texttt{crc32} &
47198 &
1842254 & (39.0x) &
36548242 & (774.4x) &
46386 &
2035556 & (43.9x) &
37694482 & (812.6x) \\
\texttt{ctl-stack} &
37440 &
616068 & (16.5x) &
14702668 & (392.7x) &
37847 &
725898 & (19.2x) &
21964505 & (580.3x) \\
\texttt{ctl-string} &
31313 &
853770 & (27.3x) &
18846973 & (601.9x) &
31845 &
704006 & (22.1x) &
19457277 & (611.0x) \\
\texttt{ctl-vector} &
30486 &
436212 & (14.3x) &
9325473 & (305.9x) &
30914 &
442518 & (14.3x) &
10432897 & (337.5x) \\
\texttt{cubic} &
33872 &
209368 & (6.2x) &
9250622 & (273.1x) &
30295 &
215268 & (7.1x) &
9827263 & (324.4x) \\
\texttt{dijkstra} &
1970036 &
44216908 & (22.4x) &
1253757348 & (636.4x) &
1961091 &
42750800 & (21.8x) &
1470626853 & (749.9x) \\
\texttt{dtoa} &
13163 &
88265 & (6.7x) &
2015472 & (153.1x) &
13020 &
78937 & (6.1x) &
2348685 & (180.4x) \\
\texttt{duff} &
6198 &
73171 & (11.8x) &
2860949 & (461.6x) &
6043 &
63287 & (10.5x) &
2248195 & (372.0x) \\
\texttt{edn} &
194838 &
3877224 & (19.9x) &
86990331 & (446.5x) &
196602 &
4120570 & (21.0x) &
104584239 & (532.0x) \\
\texttt{expint} &
29623 &
468879 & (15.8x) &
4908536 & (165.7x) &
29713 &
453217 & (15.3x) &
5642536 & (189.9x) \\
\texttt{fac} &
4334 &
77443 & (17.9x) &
1858161 & (428.7x) &
4226 &
62654 & (14.8x) &
2150388 & (508.8x) \\
\texttt{fasta} &
2241455 &
36760694 & (16.4x) &
718824003 & (320.7x) &
2285382 &
38691950 & (16.9x) &
770811211 & (337.3x) \\
\texttt{fdct} &
7519 &
25085 & (3.3x) &
1528669 & (203.3x) &
7465 &
38172 & (5.1x) &
2103383 & (281.7x) \\
\texttt{fibcall} &
2713 &
20411 & (7.5x) &
827897 & (305.2x) &
2731 &
22243 & (8.1x) &
893172 & (327.0x) \\
\texttt{fir} &
772946 &
17543810 & (22.7x) &
804645036 & (1041.0x) &
855949 &
22698227 & (26.5x) &
827569902 & (966.8x) \\
\texttt{frac} &
12313 &
172861 & (14.0x) &
5878181 & (477.4x) &
12136 &
185299 & (15.3x) &
5985378 & (493.2x) \\
\texttt{huffbench} &
1417998 &
14160500 & (10.0x) &
216244934 & (152.5x) &
1449947 &
13955966 & (9.6x) &
229499477 & (158.3x) \\
\texttt{insertsort} &
4212 &
81626 & (19.4x) &
2250385 & (534.2x) &
4065 &
95022 & (23.4x) &
2758031 & (678.3x) \\
\texttt{janne\_complex} &
2272 &
17739 & (7.8x) &
449771 & (197.9x) &
2171 &
18005 & (8.3x) &
484653 & (223.2x) \\
\texttt{jfdctint} &
11636 &
142194 & (12.2x) &
2969790 & (255.2x) &
11497 &
123960 & (10.8x) &
3773542 & (328.2x) \\
\texttt{lcdnum} &
2065 &
22535 & (10.9x) &
323111 & (156.4x) &
1950 &
16860 & (8.6x) &
395945 & (203.0x) \\
\texttt{levenshtein} &
153352 &
3282550 & (21.4x) &
108037727 & (704.5x) &
149114 &
4310603 & (28.9x) &
114304863 & (766.6x) \\
\texttt{ludcmp} &
8823 &
82479 & (9.3x) &
2434265 & (275.9x) &
8765 &
83908 & (9.6x) &
3208591 & (366.0x) \\
\texttt{matmult-float} &
68462 &
1689146 & (24.7x) &
50348408 & (735.4x) &
67895 &
1762057 & (26.0x) &
57860913 & (852.2x) \\
\texttt{matmult-int} &
443996 &
9399015 & (21.2x) &
375279612 & (845.2x) &
443126 &
11765284 & (26.6x) &
501250720 & (1131.2x) \\
\texttt{mergesort} &
525439 &
10632506 & (20.2x) &
222978650 & (424.4x) &
514642 &
10430846 & (20.3x) &
281787323 & (547.5x) \\
\texttt{miniz} &
3270 &
14691 & (4.5x) &
412024 & (126.0x) &
3143 &
13570 & (4.3x) &
487135 & (155.0x) \\
\texttt{minver} &
6194 &
53678 & (8.7x) &
1413440 & (228.2x) &
6373 &
52712 & (8.3x) &
1506640 & (236.4x) \\
\texttt{nbody} &
256466 &
6453615 & (25.2x) &
191264270 & (745.8x) &
252409 &
5888723 & (23.3x) &
216254299 & (856.8x) \\
\texttt{ndes} &
114508 &
1772845 & (15.5x) &
54788991 & (478.5x) &
119261 &
1683039 & (14.1x) &
65749412 & (551.3x) \\
\texttt{nettle-aes} &
114333 &
567288 & (5.0x) &
31052424 & (271.6x) &
112622 &
576548 & (5.1x) &
34106112 & (302.8x) \\
\texttt{nettle-arcfour} &
88005 &
2120577 & (24.1x) &
26954748 & (306.3x) &
86661 &
1436640 & (16.6x) &
35980953 & (415.2x) \\
\texttt{nettle-cast128} &
13974 &
26723 & (1.9x) &
206736 & (14.8x) &
12886 &
23546 & (1.8x) &
317716 & (24.7x) \\
\texttt{nettle-des} &
9078 &
32316 & (3.6x) &
255601 & (28.2x) &
8921 &
23980 & (2.7x) &
238207 & (26.7x) \\
\texttt{nettle-md5} &
7350 &
32253 & (4.4x) &
675813 & (91.9x) &
7217 &
29839 & (4.1x) &
937431 & (129.9x) \\
\texttt{nettle-sha256} &
14889 &
65280 & (4.4x) &
1735502 & (116.6x) &
14623 &
62231 & (4.3x) &
1652764 & (113.0x) \\
\texttt{newlib-exp} &
3658 &
25479 & (7.0x) &
661410 & (180.8x) &
3352 &
18812 & (5.6x) &
639414 & (190.7x) \\
\texttt{newlib-log} &
3038 &
16828 & (5.5x) &
511174 & (168.2x) &
2908 &
12749 & (4.4x) &
546217 & (187.8x) \\
\texttt{newlib-mod} &
2098 &
12596 & (6.0x) &
286197 & (136.4x) &
2010 &
13121 & (6.5x) &
337327 & (167.8x) \\
\texttt{newlib-sqrt} &
9430 &
186240 & (19.7x) &
6294054 & (667.4x) &
9174 &
183378 & (20.0x) &
5230626 & (570.1x) \\
\texttt{ns} &
28058 &
823601 & (29.4x) &
32090464 & (1143.7x) &
23796 &
943435 & (39.6x) &
32109538 & (1349.3x) \\
\texttt{nsichneu} &
12323 &
18872 & (1.5x) &
160230 & (13.0x) &
12236 &
18332 & (1.5x) &
164601 & (13.5x) \\
\texttt{prime} &
58925 &
944634 & (16.0x) &
16115583 & (273.5x) &
57233 &
1005099 & (17.6x) &
18090661 & (316.1x) \\
\texttt{qsort} &
3625 &
34736 & (9.6x) &
810665 & (223.6x) &
3543 &
28630 & (8.1x) &
1046006 & (295.2x) \\
\texttt{qurt} &
5333 &
67598 & (12.7x) &
1442175 & (270.4x) &
5140 &
64387 & (12.5x) &
1418908 & (276.0x) \\
\texttt{recursion} &
7838 &
177321 & (22.6x) &
5943982 & (758.3x) &
6813 &
147406 & (21.6x) &
5670809 & (832.3x) \\
\texttt{rijndael} &
1846763 &
10954085 & (5.9x) &
298854160 & (161.8x) &
1829801 &
11877918 & (6.5x) &
301644496 & (164.9x) \\
\texttt{select} &
2304 &
17312 & (7.5x) &
335812 & (145.7x) &
2182 &
18485 & (8.5x) &
369037 & (169.1x) \\
\texttt{sglib-arraybinsearch} &
32531 &
1061435 & (32.6x) &
28361404 & (871.8x) &
31769 &
1107230 & (34.9x) &
31999324 & (1007.2x) \\
\texttt{sglib-arrayheapsort} &
74568 &
1174730 & (15.8x) &
41340255 & (554.4x) &
73905 &
1196702 & (16.2x) &
50085952 & (677.7x) \\
\texttt{sglib-arrayquicksort} &
36200 &
583738 & (16.1x) &
26300739 & (726.5x) &
35305 &
545617 & (15.5x) &
25654013 & (726.6x) \\
\texttt{sglib-dllist} &
104108 &
947183 & (9.1x) &
26737508 & (256.8x) &
102530 &
930507 & (9.1x) &
31315926 & (305.4x) \\
\texttt{sglib-hashtable} &
72956 &
752795 & (10.3x) &
23759580 & (325.7x) &
74857 &
798844 & (10.7x) &
27269711 & (364.3x) \\
\texttt{sglib-listinsertsort} &
150571 &
3901838 & (25.9x) &
79469309 & (527.8x) &
151002 &
5126991 & (34.0x) &
99854495 & (661.3x) \\
\texttt{sglib-listsort} &
83492 &
865237 & (10.4x) &
26506487 & (317.5x) &
83432 &
879144 & (10.5x) &
31338183 & (375.6x) \\
\texttt{sglib-queue} &
79923 &
1171442 & (14.7x) &
33365763 & (417.5x) &
79308 &
1213322 & (15.3x) &
41832831 & (527.5x) \\
\texttt{sglib-rbtree} &
197550 &
1975689 & (10.0x) &
55392435 & (280.4x) &
198785 &
1797773 & (9.0x) &
59307610 & (298.3x) \\
\texttt{slre} &
81343 &
966902 & (11.9x) &
30416350 & (373.9x) &
79819 &
1022986 & (12.8x) &
34449484 & (431.6x) \\
\texttt{sqrt} &
281429 &
6231013 & (22.1x) &
91157327 & (323.9x) &
286888 &
4148972 & (14.5x) &
87278192 & (304.2x) \\
\texttt{st} &
46787 &
879737 & (18.8x) &
13216784 & (282.5x) &
45113 &
591158 & (13.1x) &
16071550 & (356.2x) \\
\texttt{statemate} &
5426 &
48227 & (8.9x) &
1187334 & (218.8x) &
5319 &
41665 & (7.8x) &
1302996 & (244.9x) \\
\texttt{stb\_perlin} &
170913 &
2951012 & (17.3x) &
81001658 & (473.9x) &
141537 &
3078432 & (21.7x) &
85481578 & (604.0x) \\
\texttt{stringsearch1} &
15688 &
95590 & (6.1x) &
5010590 & (319.4x) &
15624 &
116866 & (7.5x) &
6019351 & (385.3x) \\
\texttt{strstr} &
5072 &
59257 & (11.7x) &
1662703 & (327.8x) &
4943 &
51456 & (10.4x) &
2064634 & (417.6x) \\
\texttt{tarai} &
2698 &
25798 & (9.6x) &
898884 & (333.1x) &
2566 &
24645 & (9.6x) &
937342 & (365.3x) \\
\texttt{trio-snprintf} &
16752 &
89957 & (5.4x) &
1578752 & (94.2x) &
16670 &
80103 & (4.8x) &
1886009 & (113.1x) \\
\texttt{trio-sscanf} &
21441 &
170602 & (8.0x) &
4218744 & (196.8x) &
21561 &
121913 & (5.7x) &
4357713 & (202.1x) \\
\texttt{ud} &
11003 &
70482 & (6.4x) &
2121308 & (192.8x) &
10692 &
77272 & (7.2x) &
2755305 & (257.7x) \\
\texttt{whetstone} &
337043 &
3331038 & (9.9x) &
87417253 & (259.4x) &
301729 &
3001540 & (9.9x) &
99761886 & (330.6x) \\
 \hline
\end{tabular}
}
\end{table*}

\begin{table*}
\caption{PARSEC performance degradation results (cycles, thousands) on Skylake and Kaby Lake.}
\label{tab:parsec1}
\centering
\resizebox{0.85\textwidth}{!}{%
\begin{tabular}{|l|rrrrr|rrrrr|} \hline
& \multicolumn{5}{|c|}{Skylake} & \multicolumn{5}{|c|}{Kaby Lake} \\
Benchmark &
\NO &
\multicolumn{2}{c}{\DG} &
\multicolumn{2}{c|}{\HD} &
\NO &
\multicolumn{2}{c}{\DG} &
\multicolumn{2}{c|}{\HD} \\ \hline
\texttt{blackscholes} &
1054281 &
1780761 & (1.7x) &
18930407 & (18.0x) &
1076981 &
1865230 & (1.7x) &
23938125 & (22.2x) \\
\texttt{streamcluster} &
1615085 &
5743487 & (3.6x) &
135072031 & (83.6x) &
1609993 &
5117245 & (3.2x) &
149581491 & (92.9x) \\
\texttt{fluidanimate} &
1811088 &
8432568 & (4.7x) &
152698071 & (84.3x) &
1819377 &
8398743 & (4.6x) &
176033235 & (96.8x) \\
\texttt{swaptions} &
1530971 &
5362703 & (3.5x) &
85922802 & (56.1x) &
1534618 &
5586172 & (3.6x) &
112188654 & (73.1x) \\
\texttt{freqmine} &
2287791 &
4632607 & (2.0x) &
59327806 & (25.9x) &
2312730 &
4426167 & (1.9x) &
70071636 & (30.3x) \\
\texttt{canneal} &
2682216 &
9233551 & (3.4x) &
104439434 & (38.9x) &
3017818 &
8427129 & (2.8x) &
123489532 & (40.9x) \\
 \hline
\end{tabular}
}
\end{table*}

\begin{table*}
\caption{PARSEC performance degradation results (cycles, thousands) on Coffee Lake and Whiskey Lake.}
\label{tab:parsec2}
\centering
\resizebox{0.85\textwidth}{!}{%
\begin{tabular}{|l|rrrrr|rrrrr|} \hline
& \multicolumn{5}{|c|}{Coffee Lake} & \multicolumn{5}{|c|}{Whiskey Lake} \\
Benchmark &
\NO &
\multicolumn{2}{c}{\DG} &
\multicolumn{2}{c|}{\HD} &
\NO &
\multicolumn{2}{c}{\DG} &
\multicolumn{2}{c|}{\HD} \\ \hline
\texttt{blackscholes} &
960317 &
1653042 & (1.7x) &
23307722 & (24.3x) &
897093 &
1728380 & (1.9x) &
35757406 & (39.9x) \\
\texttt{streamcluster} &
1612074 &
5753689 & (3.6x) &
150934680 & (93.6x) &
1438805 &
5510738 & (3.8x) &
263693538 & (183.3x) \\
\texttt{fluidanimate} &
1616088 &
7346424 & (4.5x) &
172996775 & (107.0x) &
1626180 &
9221194 & (5.7x) &
212095965 & (130.4x) \\
\texttt{swaptions} &
1614836 &
5397065 & (3.3x) &
92169463 & (57.1x) &
1356684 &
6204397 & (4.6x) &
149384195 & (110.1x) \\
\texttt{freqmine} &
2152394 &
4007422 & (1.9x) &
69069190 & (32.1x) &
2113832 &
4481302 & (2.1x) &
80329274 & (38.0x) \\
\texttt{canneal} &
2567086 &
7895577 & (3.1x) &
112026498 & (43.6x) &
2566824 &
9135742 & (3.6x) &
127757591 & (49.8x) \\
 \hline
\end{tabular}
}
\end{table*}

\bibliographystyle{plainnat}%

\end{document}